\documentclass[aps,reprint,twocolumn,amsmath,amssymb,longbibliography,superscriptaddress]{revtex4-1}
\pdfpageattr {/Group << /S /Transparency /I true /CS /DeviceRGB>>}
\usepackage{graphicx}
\usepackage{color}
\usepackage[hidelinks]{hyperref}
\usepackage{orcidlink}
\usepackage{amsmath}
\usepackage{amsthm}
\usepackage{physics}
\usepackage{amsfonts}
\usepackage{times,txfonts}
\usepackage{bm}
\usepackage{xcolor}
\usepackage{braket,dsfont}

\hypersetup{colorlinks=true, linkcolor=blue, citecolor=blue, urlcolor=blue}

\begin{document}

\author{George Mihailescu\,\orcidlink{0000-0002-0048-9622}}
 \email[]{george.mihailescu@ucdconnect.ie}
 \affiliation{School of Physics, University College Dublin, Belfield, Dublin 4, Ireland}
 \affiliation{Centre for Quantum Engineering, Science, and Technology, University College Dublin, Dublin 4, Ireland}

\author{Saubhik Sarkar\,\orcidlink{0000-0002-2933-2792}}
\affiliation{Institute of Fundamental and Frontier Sciences, University of Electronic Science and Technology of China, Chengdu 611731, China}
\affiliation{Key Laboratory of Quantum Physics and Photonic Quantum Information, Ministry of Education, University of Electronic Science and Technology of China, Chengdu 611731, China}

\author{Abolfazl Bayat\,\orcidlink{0000-0003-3852-4558}}
\affiliation{Institute of Fundamental and Frontier Sciences, University of Electronic Science and Technology of China, Chengdu 611731, China}
\affiliation{Key Laboratory of Quantum Physics and Photonic Quantum Information, Ministry of Education, University of Electronic Science and Technology of China, Chengdu 611731, China}

 \author{Steve Campbell\,\orcidlink{0000-0002-3427-9113}}
 \affiliation{School of Physics, University College Dublin, Belfield, Dublin 4, Ireland}
 \affiliation{Centre for Quantum Engineering, Science, and Technology, University College Dublin, Dublin 4, Ireland}

 \author{Andrew K. Mitchell\,\orcidlink{0000-0002-0652-2710}}
 \email[]{andrew.mitchell@ucd.ie}
 \affiliation{School of Physics, University College Dublin, Belfield, Dublin 4, Ireland}
 \affiliation{Centre for Quantum Engineering, Science, and Technology, University College Dublin, Dublin 4, Ireland}

\title{Metrological symmetries in singular quantum multi-parameter estimation}


\begin{abstract}
\noindent The theoretical foundation of quantum sensing is rooted in the Cram\'{e}r-Rao formalism, which establishes quantitative precision bounds for a given quantum probe. In many practical scenarios, where more than one parameter is unknown, the \textit{multi-parameter} Cram\'{e}r-Rao bound (CRB) applies. Since this is a matrix inequality involving the inverse of the quantum Fisher information matrix (QFIM), the formalism breaks down when the QFIM is singular. In this paper, we examine the physical origins of such singularities, showing that they result from an over-parametrization on the metrological level. This is itself caused by emergent \textit{metrological symmetries}, whereby the same set of measurement outcomes are obtained for different combinations of system parameters. Although the number of effective parameters is equal to the number of non-zero QFIM eigenvalues, the Cram\'{e}r-Rao formalism typically does not provide information about the effective parameter encoding. Instead, we demonstrate through a series of concrete examples that Bayesian estimation can provide deep insights. In particular, the metrological symmetries appear in the Bayesian posterior distribution as lines of persistent likelihood running through the space of unknown parameters. These lines are contour lines of the effective parameters which, through suitable parameter transformations, can be estimated and follow their own effective CRBs.
\end{abstract}
\maketitle

\section{Introduction}
Fundamental scientific advances are often facilitated by breakthroughs in calibration and readout techniques that increase the measurement precision of sensors. Quantum sensing is a prime example in which quantum features, such as entanglement~\cite{giovannetti2004quantum,Giovannetti2006quantum,kuriyattil2024entangledstatessparselycoupled,giovannetti2011advances}, squeezing~\cite{grangier1987squeezed,polzik1992spectroscopy,*polzik1992atomic,gietka2023squeezing} and criticality~\cite{campos2007quantum,zanardi2008quantum,rams2018limits,garbe2020critical,chu2021dynamic,sarkar2022free,di2023critical,salvia2023critical,alushi2024optimality,puig2024dynamical,mihailescu2024multiparameter}, are harnessed to enhance the precision of parameter estimation well beyond the capacity of their classical counterparts~\cite{degen2017quantum,braun2018quantum,montenegro2024review}. 
Quantum sensors have now been developed in various physical platforms such as optical setups~\cite{mitchell2004super, pezze2007phase, ono2013entanglement}, ion-trap systems~\cite{leibfried2004toward}, cold Bosonic atoms~\cite{gross2010nonlinear}, nitrogen vacancy centres~\cite{bonato2016optimized}, superconducting qubits~\cite{wang2019heisenberg, yu2025experimental}, nuclear magnetic resonance systems~\cite{liu2021experimental}, and Rydberg atomic systems~\cite{ding2022enhanced}. More recently, nanoelectronic devices have been put forward as a promising alternative~\cite{mihailescu2024quantum,petropoulos2024nanoscale}, since complex many-body states of quantum matter can be engineered in such systems \cite{barthelemy2013quantum,sen2024many}, including nontrivial critical points \cite{iftikhar2018tunable,pouse2023quantum}, that could be leveraged for sensing. Quantum sensors are now used to search for elementary particles~\cite{jiang2021search,sushkov2023quantum}, measuring fundamental constants of the universe~\cite{rosi2014precision}, developing ultra-precise clocks~\cite{huntemann2016single,king2022optical}, thermometry~\cite{mihailescu2023thermometry,deffner2024enhancingprecisionthermometrynonlinear,brattegard2024thermometry,PhysRevLett.127.190402,mok2021optimal,mehboudi2022fundamental,srivastava2023topological,correa2015individual,PhysRevA.104.052214,mehboudi2019thermometry}, biological inter-cell monitoring~\cite{aslam2023quantum} and measuring electric~\cite{ding2022enhanced,gilmore2021quantum}, magnetic~\cite{aiello2013composite,Barry2020sensitivitry,gottscholl2021spin,maletinsky2012robust,schaffner2024quantum} or gravitational fields~\cite{stray2022quantum,van2010bose,muntinga2013interferometry,braun2025metrologygravitationaleffectsmechanical,PhysRevA.111.012411} with unprecedented precision. The theoretical basis for quantum sensing is rooted in the Cram\'{e}r-Rao inequality, which establishes fundamental limits on the achievable precision of quantum sensors~\cite{braunstein1994statistical,liu2019quantum}. This framework provides a complete understanding of single-parameter sensing, delivering a tight, saturable bound on precision as well as a systematic approach for identifying the optimal measurement basis that achieves this bound, see Ref.~\cite{paris2009quantum} for a detailed review.

However in many practical situations, there is uncertainty in more than just one parameter, and then the multi-parameter estimation framework must be used instead~\cite{paris2009quantum,liu2019quantum}. This has several important consequences that must be taken into account when designing the metrological strategy. Typically the precision of quantum sensing for a given parameter is negatively affected by uncertainty in other parameters \cite{mihailescu2024uncertain}. This is still captured by the Cram\'{e}r-Rao formalism, but it becomes considerably more complex for multi-parameter sensing. The precision is quantified by a covariance matrix whose elements describe the interdependence of the estimated parameters. When the measurement setup is specified, the Cram\'{e}r-Rao formalism bounds the covariance matrix by the inverse of the classical Fisher information matrix (CFIM), $\mathcal{\hat{F}}$. The lowest possible bound is then given by optimizing over all possible measurements. The Cram\'{e}r-Rao bound (CRB) is then provided by the inverse of the \textit{quantum} Fisher information matrix (QFIM), $\mathcal{\hat{I}}$. Because the CRB is a \textit{matrix} inequality in the case of multi-parameter estimation, the bounds for different parameters are often not simultaneously saturable. This happens when the optimal measurements for different parameters do not commute. This issue, also known as measurement incompatibility, has recently been the subject of intense study~\cite{carollo2019quantumness,demkowicz2020multi,albarelli2020perspective,suzuki2020quantum,tsang2020quantum,ragy2016compatibility,gorecki2020optimal,albarelli2022probe,gorecki2022multiparameter}. 

However, another (but much less well studied) aspect of multi-parameter quantum sensing is the possibility that the Fisher information matrices may become singular -- that is, not invertible. In fact, this scenario is not uncommon~\cite{dtamin}. In such a case, the Cram\'{e}r-Rao formalism breaks down and the multi-parameter CRB, which involves the matrix inverse of the CFIM or QFIM, becomes undefined. 
The CFIM can be singular even when the QFIM is not singular. This happens when the measurement setup is incomplete~\cite{Candeloro_2024}. 
However, this problem can in principle always be overcome by simply changing the measurement basis, switching to POVMs, or exploiting sequential projective measurements~\cite{yang2025overcoming}. 

The singularity of the QFIM, however, presents a far more fundamental challenge, since this implies that the CFIM must also be singular for all possible measurement settings. We show in this paper that the QFIM becomes singular when even optimal measurements cannot distinguish between systems with different sets of the unknown parameters. The statistical models constructed using the measurement data are in such cases identical, and this ambiguity spoils the standard estimation strategy. The QFIM singularity implies a ``metrological symmetry'' in which the set of measurement outcomes are asymptotically invariant to certain transformations of system parameters. This arises due to an effective over-parametrization on the metrological level; this can happen even when the quantum \textit{state} used for estimation has no such over-parametrization. We show that the number of non-zero eigenvalues of the QFIM is equal to the number of effective parameters, whereas the zero eigenvalues correspond to constraints that give rise to the metrological symmetries. However, the effective parameter encoding and the effective CRBs for these parameters is extremely challenging to extract from analysis of a singular QFIM.

We therefore turn to Bayesian estimation, and show that not only are unambiguous signatures of QFIM singularities immediately obvious in the multinomial posterior distribution, but that detailed information about the effective parameter encoding and metrological symmetries can be straightforwardly read off. This information can be used to perform a re-parameterization (which is in general a non-linear coordinate transformation), for which the marginalized probability distributions of effective parameters tend to Gaussian and follow effective CRB scaling. The values of the effective parameters can therefore be reliably estimated within the Bayesian strategy, even for singular multi-parameter problems.

\begin{figure}[t]
    \centering
    \includegraphics[width=1\linewidth]{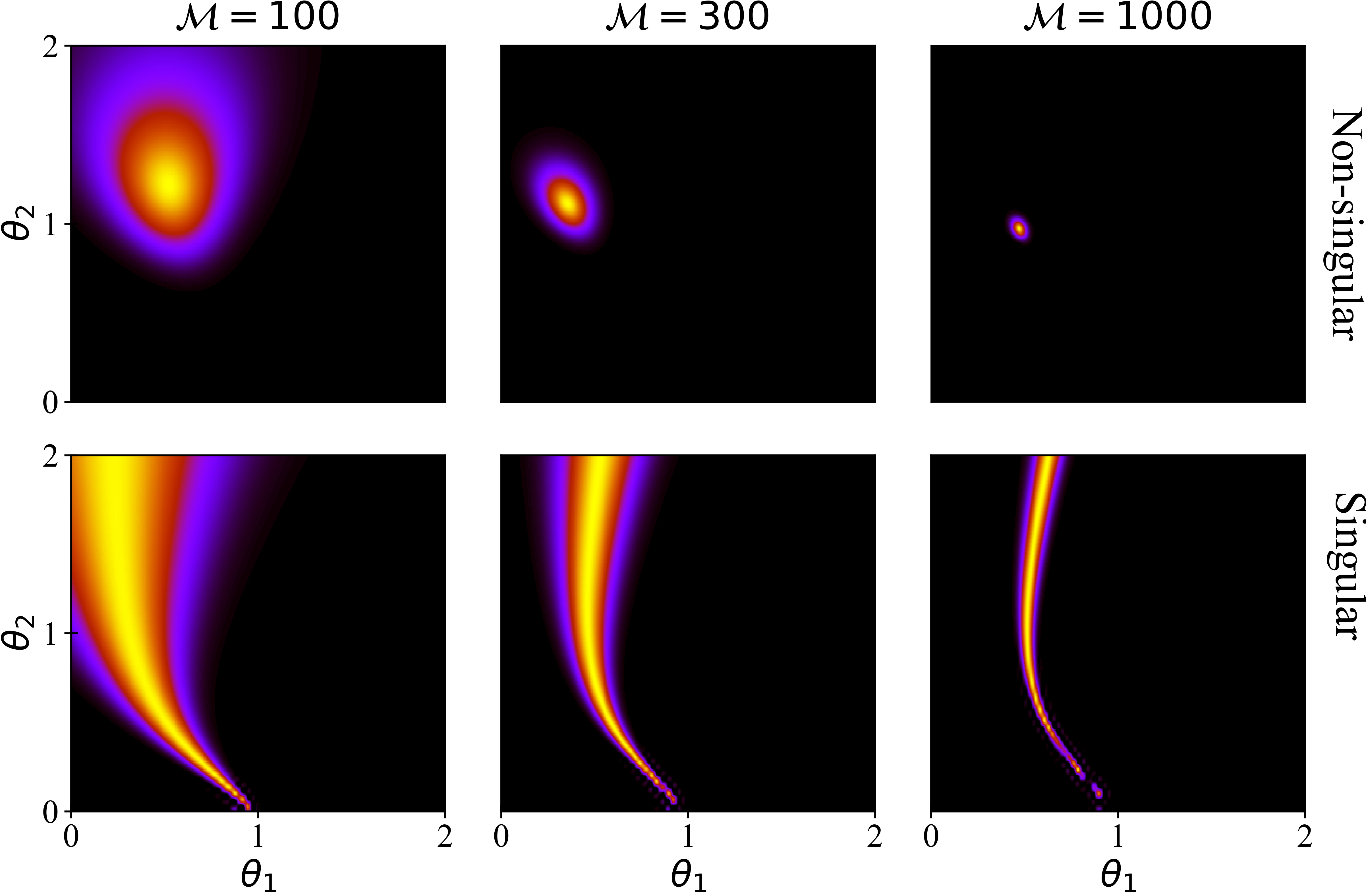}
    \caption{Illustration contrasting Bayesian estimation for singular vs non-singular multi-parameter metrological problems. We show typical posterior distributions after $\mathcal{M}=100$, $300$ and $1000$ measurements, with the non-singular case (top panels) converging to a \textit{point} in the space of unknown parameters $\theta_1$ and $\theta_2$; whereas in the singular case (bottom panels) the distribution converges instead to a \textit{line}, encoding an effective metrological symmetry. The example shown is for the Heisenberg trimer, see Appendix~\ref{app:triangle} for further details.}
    \label{fig:illustration}
\end{figure}

This is illustrated in Fig.~\ref{fig:illustration}, where we compare the Bayesian posterior distribution for estimation of two parameters $\theta_1$ and $\theta_2$ after $\mathcal{M}=100$, $300$ and $1000$ measurements (left to right) for the standard non-singular case (upper panels) and a situation where the QFIM is singular (lower panels). The specific example shown is for the Heisenberg trimer model (see Appendix~\ref{app:triangle}) but the qualitative behavior is very generic. In the non-singular case, the posterior converges to a unique \textit{point} in parameter space as more data are collected, corresponding to the true values of the unknown parameters.
By contrast, for a singular system a persistent \textit{line} of high likelihood emerges. This line encodes the metrological symmetry: the line passes through the true value of the parameters, but measurements cannot distinguish systems with different parameter values that lie along this line. The line itself sharpens up as more data are collected, and converges to a contour line of the \textit{effective} parameter. The effective parameter can therefore be estimated, with a precision that follows an effective CRB.

In this paper we examine the physical origins of QFIM singularities, discuss their connection to metrological symmetries, and show how Bayesian estimation can be used to uncover these metrological symmetries. We show through concrete examples that the functional dependence of effective parameters on the original parameters can be extracted, and that these effective parameters can be reliably estimated.

\section{Parameter estimation}

In this section, we provide a brief review of quantum estimation theory. We focus on two different techniques to analyse quantum sensing problems, the frequentist and Bayesian approaches, and show how the two are connected through the Bernstein–von Mises theorem~\cite{le1956asymptotic,le2012asymptotic}. We highlight key results that will be used throughout the rest of the paper, however readers who are already familiar with these methodologies may wish to proceed directly to Sec.~\ref{sec:ConventionalCase}.


\subsection{Frequentist approach}
The most general quantum sensing scenario arises when there is a set of $d$ unknown parameters $\vec{\theta} = \left( \theta_1, \theta_2, ..., \theta_d \right)$ to be estimated through suitable measurements of a quantum probe. Information about these parameters is imprinted on the state of the probe, described by the density matrix $\hat{\varrho} \left( \vec{\theta} \right)$. 
For a given set of POVMs $\{ \Pi_k \}$, the uncertainty of inferring the set of parameters from $\mathcal{M}$ measurements on the probe is set by the Cram\'{e}r-Rao Bound (CRB)~\cite{braunstein1994statistical, paris2009quantum, liu2019quantum},
\begin{equation}
    \label{eq:classical_crb}
    \text{Cov}\left[ \hat{\vec{\theta}} \right] \geq \frac{1}{\mathcal{M}} \hat{\mathcal{F}}^{-1}\;,
\end{equation}
where $\text{Cov}[ \hat{\vec{\theta}} ]$ represents the covariance matrix whose elements are given by $\text{Cov}( \theta_i, \theta_j){ = }\langle ( \theta_i {-} \langle \theta_i \rangle ) ( \theta_j {-} \langle \theta_j \rangle ) \rangle$ and 
$\hat{\mathcal{F}}$ is the CFIM whose elements are given by,
\begin{equation}
    \label{Eq:Multi_Param_FI_Def}
	\hat{\mathcal{F}}_{i,j} = \mathbb{E}\left[ \left( \partial_{\theta_i}\ln{p\left(x_k \mid \vec{\theta}\right)}\right)\left( \partial_{\theta_j}\ln{p\left(x_k \mid \vec{\theta}\right)}\right)\right]\;,
\end{equation}
where $p(x_k | \vec{\theta} )$ denotes the probability of obtaining measurement outcome $x_k$ conditioned on the parameters having a value $\vec{\theta}$ through the Born rule $p(x_k | \vec{\theta} ) = \Tr{\Pi_k \hat{\varrho} ( \vec{\theta} )}$.

The QFIM $\hat{\mathcal{I}}$ is obtained by maximizing over all possible POVMs. Its elements can be expressed in terms of the symmetric logarithmic derivative (SLD) operators as \cite{liu2019quantum},
\begin{equation}\label{eq:QFIM_sld}
 \hat{\mathcal{I}}_{i,j} = \tfrac{1}{2} \Tr\left( \hat{\varrho} \acomm{\hat{L}_i}{\hat{L}_j} \right) \equiv  \text{Tr}\left( \hat{L}_j \partial_{\theta_i} \hat{\varrho}(\vec{\theta})  \right) \equiv - \text{Tr}\left( \hat{\varrho}(\vec{\theta}) \partial_{\theta_i} \hat{L}_j \right) \;,
\end{equation}
where $\hat{L}_i$ the SLD operator for parameter $\theta_i$. Formally, the SLD is implicitly defined by the solution of the self-adjoint operator equation, 
\begin{equation}
    \partial_{\theta_i} \hat{\varrho}(\vec{\theta}) = \frac{1}{2}\left(\hat{L}_i \hat{\varrho}(\vec{\theta}) + \hat{\varrho}(\vec{\theta}) \hat{L}_i \right)\;.
\end{equation}
We note that diagonal elements of the QFIM are identical to the single parameter QFIs, $\hat{\mathcal{I}}_{i,i} \!\equiv\! \mathcal{I}\left(\theta_i\right)$ while the off-diagonal elements encode correlations between measurements. The QFIM is a real-symmetric, positive-semidefinite matrix. Using the spectral decomposition of the density matrix $\hat{\varrho} ( \vec{\theta} )  = \sum_i e_i \ket{e_i}\bra{e_i}$, elements of the QFIM can be written as, 
\begin{align}
    \label{eq:QFIM_Entry}
    {\mathcal{I}}_{i,j} = \sum_{k,l}\frac{2\Re\left[\langle e_k|\partial_{\theta_i\:} \hat{\varrho}| e_l \rangle\langle e_l|\partial_{\theta_j\:} \hat{\varrho} | e_k\rangle\right]}{e_k + e_l}\;,
\end{align}
which for pure states $\hat{\varrho} = \ket{\psi}\bra{\psi}$ simplifies to
\begin{align}
    \label{eq:Pure_State_QFI}
    \mathcal{I}_{i,j} = 4\Re\left[\langle \partial_{\theta_i} \psi|\partial_{\theta_j} \psi\rangle - \langle \partial_{\theta_i} \psi|\psi\rangle \langle \psi| \partial_{\theta_j} \psi \rangle \right]\;.
\end{align}

The fundamental attainable precision of multi-parameter estimation is  set by the quantum CRB,
\begin{equation}
    \label{eq:QCRB_Def}
    \text{Cov} \left[ \hat{\vec{\theta}} \right] \geq  \frac{1}{\mathcal{M}} \hat{\mathcal{F}}^{-1}  \geq  \frac{1}{\mathcal{M}} \hat{\mathcal{I}}^{-1}\;,
\end{equation}
which is again a matrix inequality. Similar to single-parameter estimation, saturation of the bound requires large datasets and assumes local knowledge as well as unbiased estimators. The SLD operator $\hat{L}_i$ corresponds to the optimal measurement to estimate parameter $\theta_i$. It is in general a complicated operator and is not necessarily practical from an experimental point of view. This can be seen from the explicit form of the SLD operators whose matrix elements read~\cite{liu2019quantum},
\begin{equation}\label{eq:sld_explicit}
\langle e_a | \hat{L}_i |e_b\rangle =\delta_{ab}\frac{\partial_{\theta_i} e_a}{e_a} - \frac{2(e_a-e_b)}{e_a+e_b}\langle e_a|\partial_{\theta_i} e_b\rangle \;.
\end{equation}
For pure states this reduces to $\hat{L}_i= 2\left( |\psi \rangle \langle \partial_{\theta_i} \psi| + |\partial_{\theta_i}\psi\rangle\langle\psi|\right)$. 
In general the bounds are not simultaneously saturable for all parameters since the SLD operators $\hat{L}_i$ for different parameters $\theta_i$ may be incompatible \cite{ragy2016compatibility,albarelli2022probe} and therefore no mutual optimal measurement basis exists. This is known as measurement incompatibility and is a key factor that potentially prevents the ultimate precision bound in Eq.~\eqref{eq:QCRB_Def} from being saturated in practice. In such scenarios, precision trade-offs are fundamentally unavoidable. The necessary and sufficient condition for attainability of the quantum multi-parameter CRB is that $\Tr\left( \hat{\varrho} [\hat{L}_i,\hat{L}_j] \right)=0$ \cite{liu2019quantum}. When this condition is not satisfied and the issue of measurement incompatibility arises, finding an optimal strategy becomes very challenging~\cite{albarelli2022probe, gorecki2022multiparameter, gorecki2022multiple}. 
We note that in these cases one may consider alternative formulations such as the Holevo Cram\'er-Rao bound~\cite{holevo2011probabilistic} or the Nagaoka-Hayashi Cram\'er-Rao bound~\cite{conlon2021efficient}. All such bounds reduce to the regular quantum multi-parameter CRB when the SLD operators commute. 

Since Eq.~\eqref{eq:QFIM_sld} features the anticommutator of SLDs, rather than the commutator, we note that even when a single optimal measurement basis exists for all parameters and the multi-parameter bound is saturated, the precision of estimation for a \textit{given} parameter is still generically \textit{lower} than that obtained when performing true single-parameter estimation, due to correlations encoded in the off-diagonal QFIM elements \cite{mihailescu2024critical}.

Apart from the measurement incompatibility issue, invertibility of the Fisher information matrices may be a problem for multi-parameter estimation. In the case of an invertible QFIM but a singular CFIM, the issue is associated to incompleteness of the chosen measurement setup~\cite{Candeloro_2024}. The solution is to extend the number of measurement outcomes through either switching from projective measurements to POVMs or by exploiting sequential projective measurements~\cite{yang2025overcoming}. The singularity of the QFIM is more dangerous as it directly implies that the CFIM is necessarily singular for \textit{all} choices of the measurement setup. Indeed, the CRB is only meaningful when the QFIM is invertible, meaning $\text{det} [ \hat{\mathcal{I}} ] > 0$. For the two-parameter estimation problem, this requires,
\begin{equation}
    \label{eq:qfim_det}
    \text{det}\left[ \hat{\mathcal{I}} \right] \equiv \mathcal{I}_{i,i} \mathcal{I}_{j,j} - \mathcal{I}_{i,j}\mathcal{I}_{j,i} > 0 \;.
\end{equation}
When $ \text{det} [ \hat{\mathcal{I}} ]\!=\!0$ the quantum CRB becomes undefined and \emph{nothing} can be said about the precision of parameter estimation~\cite{mihailescu2024critical,dtamin,frigerio2024overcomingsloppinessenhancedmetrology,yang2025overcoming,Candeloro_2024,navarro2025existenceunbiasedresilientestimators,mihailescu2024uncertain,niu2025rolelongrangeinteractioncritical,Seveso_2020} of the parameters $\vec{\theta}$ within this formalism. A singular QFIM implies that there are no unbiased estimators for $\vec{\theta}$ with finite variance~\cite{stoica2001parameter}, and therefore a vanishing signal-to-noise ratio for parameter estimation. 
In the later sections of this paper, we explore when the QFIM becomes singular and how information about the system can nevertheless still be extracted in such cases. We focus on QFIM singularities in this paper rather than the incompatibility issue. To disentangle the effects, the examples we consider have commuting SLDs and so the QFIM already provides a full characterization.


\subsection{Bayesian approach}
Bayesian estimation theory offers a useful and versatile approach to quantum metrology~\cite{gammelmark2013bayesian,kiilerich2010bayesian,alves2022bayesian,glatthard2022optimal,cimini2024benchmarking,rubio2020bayesian,DAurelio_2022,mehboudi2022fundamental}. Under standard conditions, asymptotically optimal estimators can be obtained using Bayesian estimation in the large dataset limit, i.e. when the number of measurements $\mathcal{M} \!\to\!\infty$. Unknown values of target parameters $\vec{\theta}$ are treated as well-defined stochastic variables. Information about this set of parameters is initially assumed to follow a distribution $P(\vec{\theta})$ called the \emph{prior}. If no information about the parameters is initially known, except that they fall in some range $\theta_i \in [ \theta_i^{~\text{min}} , \theta_i^{~\text{max}} ]$, then the prior can be taken to be a uniform distribution $P(\vec{\theta}) = \Pi_i (\theta_i^{~\text{max}} - \theta_i^{~\text{min}})^{-1}$ \footnote{The specific choice of prior is largely immaterial since the converged posterior distribution is a fixed point of the Bayesian update map, and our conclusions are unaffected by using other priors.}. Such a distribution indicates our limited \textit{a priori} knowledge of the parameters $\vec{\theta}$. Bayesian estimation as considered here represents a global perspective on sensing, as opposed to the local one assumed by the frequentist approach \cite{PhysRevLett.127.190402,mukhopadhyay2024currenttrendsglobalquantum,PhysRevA.110.L030401,meyer2023quantummetrologyfinitesampleregime,PhysRevResearch.6.023305,montenegro2021global,boeyens2025rolesymmetrygeometryglobal}.

For a given set of $\mathcal{M}$ measurements, each outcome $k$ appears $n_k$ times such that $\sum_k n_k=\mathcal{M}$. The distribution encoding our updated knowledge from the observed measurement data is determined by Bayes' theorem,
\begin{equation}
    \label{eq:post_def}
P \left( \vec{\theta} \mid \{ n_k \} \right) = \frac{P \left( \{ n_k \} \mid \vec{\theta} \right) P\left( \vec{\theta} \right)}{P\left( \{ n_k \} \right)}\:,
\end{equation}
where the \emph{posterior} $P \left( \vec{\theta} \mid  \{ n_k \} \right)$ determines the remaining uncertainty about the true parameter values and represents the conditional distribution of the parameters $\vec{\theta}$ given the observed measurement data $\{ n_k \}$. The denominator $P\left( \{ n_k \} \right)$ ensures normalization since the posterior should represent a valid probability distribution. The \emph{likelihood} function, 
$$P \left( \{ n_k \} \mid \vec{\theta} \right) = \frac{\mathcal{M} !}{\prod_k n_k !} \prod_k \left( p_k^{\rm th} \right)^{n_k},$$
is a multinomial distribution in the multi-parameter case, computed using the measurement data $\{ n_k \}$. It represents the conditional probability of obtaining a particular dataset, $\{ n_k \}$, given that the parameters are $\vec{\theta}$. 
Here, $n_k$ is the frequency of the specific outcome $k$, such that $p_k^{\rm exp} = n_k/\mathcal{M}$ is the experimental likelihood that a given measurement yields outcome $k$. On the other hand, $p_k^{\rm th} = \Tr \left[ \hat{\varrho} ( \vec{\theta} ) \hat{\Pi}_k \right]$ is the theoretical probability of obtaining outcome $k$ given the parameters have value $\vec{\theta}$ from a set of POVMs $\{ \hat{\Pi}_k \}$. In the limit of large $\mathcal{M}$, $p_k^{\rm exp}$ approaches the theoretical value of $p_k^{\rm th}$.  After $\mathcal{M}$ measurements, the Bayesian posterior distribution is computed via Eq.~\eqref{eq:post_def} from the prior distribution and the measurement data $\{ n_k \}$. The prior is then updated, $  P(\vec{\theta}) \mapsto P\left( \vec{\theta} \mid \{ n_k \} \right)$ and a new set of  measurements can be used to further improve our knowledge about the unknown parameters.

The Bayesian estimate $\hat{\theta}_i$ of parameter $\theta_i$ obtained at the end of this process, after $\mathcal{M}$ measurements on the system, is given by the Bayesian average,
\begin{equation}
    \label{eq:bayesian_average}
    \hat{\theta}_i = \int dV\: \theta_i P\left(\vec{\theta} \mid \{n_k\}\right) \;,
\end{equation}
where $dV=d\theta_1 d\theta_2 ... d\theta_d$ is the metrological volume element in parameter space, such that the integral runs over all unknown parameters. Furthermore, the individual posterior distribution for parameter $\theta_i$ can be obtained by  marginalizing the multinomial distribution over the other unknown parameters,
\begin{equation}
    \label{eq:individual_posterior}
    P\left(\theta_i \mid \{n_k\} \right) = \int dV_i\: P\left(\vec{\theta} \mid \{n_k\} \right) \;,
\end{equation}
where $dV_i=dV/d\theta_i$. 

For finite $\mathcal{M}$ the variance of the posterior is finite and gives a measure of the precision of the Bayesian estimator $\hat{\theta_i}$,
\begin{equation}
    \label{eq:bayesian_variance}
    \text{Var}\left(\hat{\theta}_i\right) = \int dV\:\theta_i^2 P\left(\vec{\theta} \mid \{n_k\}\right) - \left[ \int dV\: \theta_i P\left(\vec{\theta} \mid \{n_k\}\right) \right]^2\;.
\end{equation}
The variance of the Bayesian estimator obtained in this way is related to the CRB, as discussed in the following subsection.


\subsection{Marrying frequentist with Bayesian:~~~~~\\Bernstein–von Mises theorem} 
For non-singular quantum multi-parameter estimation problems, the Bernstein–von Mises theorem (BVM) theorem~\cite{le1956asymptotic,le2012asymptotic} ensures, through the central-limit theorem, that the Bayesian posterior distribution Eq.~\eqref{eq:post_def} converges to a multivariate Gaussian distribution, centered on the true value of the parameters, $\vec{\theta}^{\rm \:tr}$. 
The width of the Gaussian reduces as the number of measurements $\mathcal{M}$ is increased. In fact, for large $\mathcal{M}$, the covariance of the posterior Gaussian is given by the quantum multi-parameter CRB \cite{le1956asymptotic,le2012asymptotic}, ${\rm Cov}[\hat{\vec{\theta}}]=\left[ \mathcal{M} \hat{\mathcal{I}}\right]^{-1}$. 
The Bayesian strategy therefore allows the unknown parameters to be extracted precisely and unambiguously in the asymptotic limit, $\mathcal{M}\to \infty$. Specifically, for large $\mathcal{M}$ we have,
\begin{equation}
    \label{eq:bvm_mp}
    P\left(\vec{\theta} \mid \{ n_k \} \right) \simeq \sqrt{ \frac{\mathcal{M}^d \text{det}\left[\hat{\mathcal{I}}(\vec{\theta}^{\rm \:tr})\right] }{(2 \pi)^d} } \exp{-\frac{\mathcal{M} (\vec{\theta} - \vec{\theta}^{\rm \:tr})^T \hat{\mathcal{I}} (\vec{\theta} - \vec{\theta}^{\rm \:tr})}{2}}\;.
\end{equation}
The BVM therefore provides the fundamental connection between the Bayesian and frequentist approaches. 

Crucially, the BVM assumes an invertible QFIM, and is inapplicable for singular estimation problems. In the singular case, the Bayesian posterior distribution generally will not converge to Eq.~\eqref{eq:bvm_mp}. Singularities in the QFIM therefore have deep implications for parameter estimation also in the Bayesian scenario. However this raises the obvious question: what does the Bayesian posterior converge to (if it does converge) for problems with a singular QFIM, and what (if anything) can be learned about the parameters to be estimated from this? These questions are answered in the following. 


\section{Multi-parameter estimation in\\ the conventional (non-singular) case}
\label{sec:ConventionalCase}

Before discussing singular cases and to contextualise our results, here we review the typically-encountered scenario of multi-parameter sensing in which the QFIM is invertible and thus conventional sensing strategies are applicable. We demonstrate this in the concrete setting of the $XY$ model,
\begin{equation}
    \label{eq:xy_general_ham}
    \hat{H}_{XY} = \sum_{i , j = 1}^{N} \frac{\lambda_{ij}}{2} \left(\left[1 + \gamma_{ij}\right]\hat{\sigma}_x^i \hat{\sigma}_x^{j} + \left[1 - \gamma_{ij}\right] \hat{\sigma}_y^i \hat{\sigma}_y^{j}  \right) + \sum_{i = 1}^{N} h \hat{\sigma}_z^i \;,
\end{equation}
which describes a system of $N$ spins-$\tfrac{1}{2}$, with arbitrary couplings $\lambda_{ij}$ between spins $i$ and $j$, and spin anisotropy $\gamma_{ij}$, subject to a magnetic field $h$ along the $z$-axis. Here $\hat{\sigma}_{x,y,z}$ denote the usual Pauli operators. 

In this section we will study $\hat{H}_{XY}$ in a ring geometry, corresponding to uniform nearest-neighbour interactions $\lambda_{ij} \equiv \lambda$ and $\gamma_{ij} \equiv \gamma$ for adjacent spins. The system can be mapped to a non-interacting fermionic model via the Jordan-Wigner transformation, and solved exactly~\cite{pfeuty1970the, barouch1970statistical,10.21468/SciPostPhysLectNotes.82} by Fourier transform (see Appendix~\ref{app:xydet}). 
We consider the (pure) ground state sensitivity to parameter variations to assess metrological utility in the multi-parameter case, employing Eq.~\eqref{eq:Pure_State_QFI} to compute the QFIM. We illustrate our results for system size $N \!=\! 4$, but remark that the qualitative behaviour is generic and applies for any $N\ge 4$. For $N<4$, the fully-connected geometry leads to a singularity of the QFIM, as discussed later in  Sec.~\ref{sec:hidden}.

\begin{figure}[t]
    \centering
    \includegraphics[width=1\linewidth]{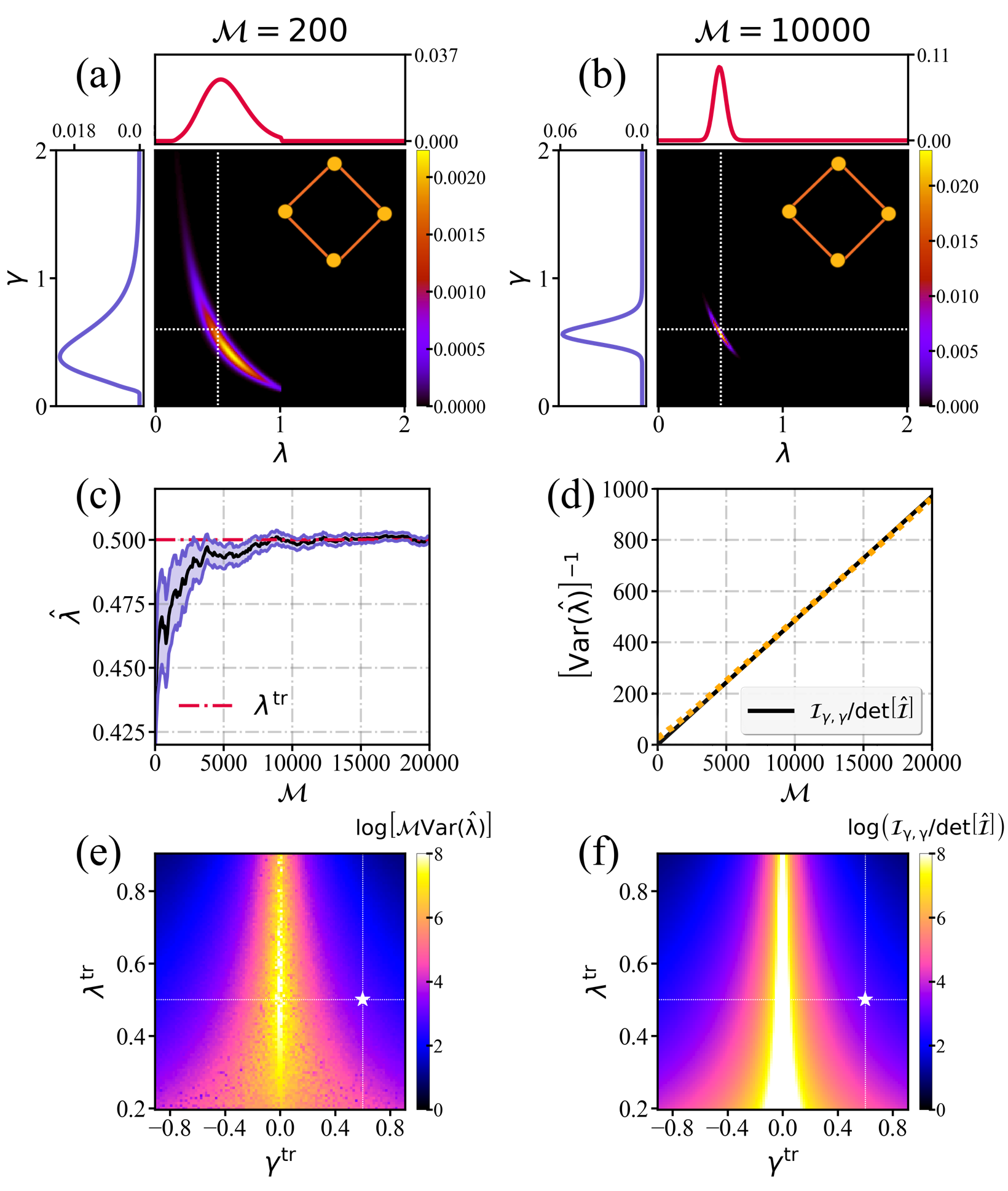}
    \caption{\textbf{Multi-parameter Bayesian estimation for the non-singular case}. We consider joint estimation of $\lambda$ and $\gamma$ using the ground state of the $XY$ model on the ring geometry with $N=4$ sites, assuming that the field $h=1$ is known. 
    Multinomial posterior distributions after $\mathcal{M} = 200$ and $10000$ measurements are shown in (a,b), with the white dotted lines denoting the true parameter values at $\left( \lambda^{\rm tr}, \gamma^{\rm tr} \right) = \left( 0.5,0.6\right)$. Line-plots show the corresponding marginalized univariate distributions for each parameter. 
    (c) Average estimate $\hat{\lambda}$ (black line) and its error (blue shaded region) as a function of $\mathcal{M}$, averaged over $500$ numerical experiments. Red line denotes the true value $\lambda^{\rm tr} = 0.5$, to which the estimator correctly converges. (d) Inverse variance of the posterior distribution for $\hat{\lambda}$ (black line) vs $\mathcal{M}$ compared with the scaling prediction from the CRB Eq.~\eqref{eq:QCRB_Def} (orange dashed line) using the QFIM from Eq.~\eqref{eq:xy_gam_lam_qfim}. (e) Full sensitivity phase diagram of posterior variance $\mathcal{M}\text{Var}\left[ \hat{\lambda} \right]$ vs $\lambda^{\rm tr}$ and $\gamma^{\rm tr}$. Results shown for $\mathcal{M} = 10000$. (f) Analogous sensitivity phase diagram obtained  from the CRB. The star point corresponds to the values used in (a-d).}
    \label{fig:xy_estimator_properties}
\end{figure}

Here we consider the dual estimation of the coupling $\lambda$ and anisotropy $\gamma$ in the $XY$-model on a ring, assuming that the field $h$ is known. The QFIM for arbitrary $N$ follows as \cite{mihailescu2024uncertain},
\begin{equation}
    \label{eq:xy_gam_lam_qfim}
    \hat{\mathcal{I}}_{\lambda, \gamma}^{\rm ring} = \sum_{k}\frac{16\sin^2(k)}{\epsilon_k^4}
    \begin{pmatrix}
    h^2 \gamma^2\;\; & h \gamma \lambda \left( h + \lambda \cos{k} \right)  \\ h \gamma \lambda \left( h + \lambda \cos{k} \right)\;\; &  \lambda^2\left(h + \lambda \cos{k}\right)^2
\end{pmatrix}\;,
\end{equation}
with momentum $k =  \pi(2n+1)/N$ for $n = 0,1,2,\ldots,\left\lfloor\frac{N}{2}\right\rfloor - 1$,  and dispersion $\epsilon_k = 2 \sqrt{(h + \lambda \cos{k} )^2 + \lambda^2\gamma^2 \sin^2{k}}$. From this we see that ${\rm det}[\hat{\mathcal{I}}_{\gamma,\lambda}^{\rm ring}] >0 $ for all $N \!\geq\! 4$ when $\gamma$, $\lambda$, and $h$ are nonzero. The invertibility of the QFIM in the ring geometry implies that the parameters $\lambda$ and $\gamma$ are estimable by making measurements on the system, with the optimal precision following the CRB, Eq.~\eqref{eq:QCRB_Def}. The finite off-diagonal elements of the QFIM indicate correlations between the measurement outcome statistics for these unknown parameters.

To showcase the success of standard Bayesian estimation for this non-singular problem, we present numerical results in Fig.~\ref{fig:xy_estimator_properties}. Here we take a fixed known field $h=1$ and seek to estimate the value of the parameters $\vec{\theta}\equiv (\lambda,\gamma)$. The true value of these parameters is taken to be $\lambda^{\rm tr}=0.5$ and $\gamma^{\rm tr}=0.6$, arbitrarily chosen. Our observable of choice is the total magnetization, 
$\hat{\sigma}_z^{\text{tot}} = \sum_i \hat{\sigma}_z^i$,
and POVMs are constructed from its eigenbasis. We motivate this choice of measurement by noting that (i) for the $N \!=\! 4$ setting considered, the total magnetization constitutes an optimal measurement, i.e.~$\comm{\hat{L}_{\theta}}{\hat{\sigma}_z^{\text{tot}}} \!=\! 0$ with $\theta$ being either $\lambda$ or $\gamma$; and (ii) this observable is an experimentally feasible measurement. The SLDs for the two parameters are compatible, which means that the corresponding CRBs for each parameter can be simultaneously saturated by measuring the magnetization. We take our initial prior to be uniform, and sample over the range $0\le \theta \le 2$ for $\theta=\lambda$ or $\gamma$. The prior is updated after ever $100$ measurements; the total number of measurements is $\mathcal{M}$. Experimental data are simulated as a series of randomly-generated measurement outcomes.

Figure~\ref{fig:xy_estimator_properties}(a,b) shows the Bayesian multinomial posterior distribution after $\mathcal{M}=200$ and $10000$ measurements, with the associated line plots being the marginalized univariate distributions for each parameter following Eq.~\eqref{eq:individual_posterior}. Importantly, as the number of measurements $\mathcal{M}$ is increased, the posterior converges to a unique \textit{point} in parameter space, corresponding to the true values of the parameters to be estimated $\vec{\theta}^{\:\rm tr}$, denoted by the white dotted lines.
This is the behavior expected from the BVM theorem for a non-singular system, with the posterior converging to the multivariate Gaussian in Eq.~\eqref{eq:bvm_mp}. The marginalized distributions are also seen to converge to univariate Gaussian profiles, with a width that decreases as $\mathcal{M}$ increases. As expected, the precision of parameter estimation improves as more data are collected.

In Fig.~\ref{fig:xy_estimator_properties}(c) the estimated value (Bayesian average) of the coupling $\hat{\lambda}$ is plotted (black line) as a function $\mathcal{M}$. This is obtained from the full posterior distribution at a given $\mathcal{M}$ via Eq.~\eqref{eq:bayesian_average}. Also shown as the blue shaded region is the error (signal noise), quantified by the standard deviation of the posterior. We see that $\hat{\lambda}$ converges to the true value $\lambda^{\rm tr}$ at large $\mathcal{M}$ and the error decreases. Furthermore, in panel (d) we plot the (inverse) variance from Eq.~\eqref{eq:bayesian_variance} (solid line), comparing with the prediction from the CRB, Eq.~\eqref{eq:QCRB_Def}, computed using the QFIM from Eq.~\eqref{eq:xy_gam_lam_qfim}. The overall linear scaling of the Bayesian estimate precision $[{\rm Var}(\hat{\lambda})]^{-1}$, as well as the precise rate of precision gain with $\mathcal{M}$, is in exact agreement with the CRB. This vividly demonstrates the agreement between the frequentist and Bayesian approaches in the limit of large $\mathcal{M}$.

One can repeat this numerical experiment for each combination of the true parameters $\vec{\theta}^{\rm \: tr}\!=\!(\lambda^{\rm tr},\gamma^{\rm tr})$, each time performing Bayesian estimation for $\lambda$ starting from a flat prior, and computing the resulting variance of the estimators after $\mathcal{M}$ measurements. For each point in the parameter space, $(\lambda^{\rm tr},\gamma^{\rm tr})$, we then compare this variance with that predicted from the CRB via the QFIM for the same $\mathcal{M}$. The resulting `sensitivity phase diagrams' are shown in Figs.~\ref{fig:xy_estimator_properties}(e,f) -- Bayesian estimate on the left, CRB prediction on the right. For $\mathcal{M}=10000$ measurements as shown, we see excellent quantitative agreement across the majority of the parameter space. Here, black and blue regions denote areas of high parameter sensitivity, i.e. low variance, whereas a yellow colour denotes a low sensitivity, i.e. large variance (note also the log-scale). Interestingly, the agreement between the CRB prediction and the Bayesian estimate deteriorates near $\gamma \!=\! 0$ and at smaller values of $\lambda$. This is, in fact, expected because the precision of estimation for the parameter $\lambda$ given by $[{\rm Var}(\hat{\lambda})]^{-1}$ is proportional to the QFIM determinant within the frequentist approach and the QFIM determinant in this case  vanishes as $\gamma,\lambda \to 0$, see Appendix \ref{app:xydet}. Not only is the sensitivity lower in this region, but Bayesian estimation becomes more challenging. For any finite QFIM determinant, the Bayesian estimate must converge with a precision given by the CRB; however the BVM theorem \cite{le1956asymptotic,le2012asymptotic} applies only asymptotically in the limit of large $\mathcal{M}$. Our results show that when the QFIM determinant is finite but small, a larger number of measurements $\mathcal{M}$ is needed in practice to attain the precision bound set by the CRB. Indeed, in regions of the phase diagram where the QFIM determinant vanishes, the inverse cannot be computed; then the CRB and the BVM theorem no longer apply.


\section{QFIM singularities: metrological symmetries and zero eigenvalues}\label{sec:diag}
In quantum sensing, measurement data are used to construct statistical models from which the values of unknown parameters are inferred. One might suppose therefore that if different sets of parameters asymptotically yield the same statistical model, then this ambiguity could lead to a breakdown in our ability to reliably estimate parameters from these measurements. From this one might guess that when measurements cannot distinguish between different sets of parameters $\vec{\theta}$ and $\vec{\theta}'$, the QFIM becomes singular. This picture implies a kind of `metrological symmetry' in singular systems, where the metrological outcome is invariant to the transformation $\vec{\theta} \to \vec{\theta}'$.  We therefore have a redundancy in the parameter encoding, leading to over-parametrization on the metrological level. In such instances the number of \emph{effective parameters} is less than the number of original parameters.

A singular QFIM has zero determinant, and therefore at least one eigenvalue must be zero, since the determinant is the product of the eigenvalues. In fact, whenever the number of effective parameters is less than the number of bare parameters, at least one zero eigenvalue is guaranteed, and therefore the QFIM is indeed necessarily singular. This makes the connection between metrological over-parametrization and QFIM singularity precise.

As we show below, \textit{the number of non-zero QFIM eigenvalues is equal to the number of effective parameters.}

This can be seen from the following argument. Consider a system with $d$ bare parameters $\vec{\theta}$ to be estimated. Since the  QFIM of these bare parameters is a real symmetric semi-definite matrix, it can always be diagonalized, 
\begin{equation}\label{eq:I_diagU}
\hat{\mathcal{D}}=\hat{U}^{T}\:\hat{\mathcal{I}}(\vec{\theta})\:\hat{U}
\end{equation} 
where $\hat{U}$ is a $d\times d$ orthogonal matrix and $\hat{\mathcal{D}}={\rm diag}(\vec{G})$ is a diagonal matrix with $k$ non-zero elements. 
Eigenvalue $G_a$ is given by 
$G_a=\sum_{i,j} U_{i, a}U_{j, a}\times \hat{\mathcal{I}}_{\theta_i,\theta_j}$ and has corresponding eigenvector $\vec{u}_a$.   
Since the QFIM $\hat{\mathcal{I}}(\vec{\theta})$ and the transformation matrix $\hat{U}(\vec{\theta})$ are both functions of the underlying parameters $\vec{\theta}$, each zero eigenvalue $G_a=0$ constitutes a \textit{constraint} of the type $h_a(\vec{\theta})=0$. Since $\hat{U}$ is orthogonal, we have $d-k$ such \textit{independent} constraints when we have $k$ non-zero eigenvalues. By the implicit inverse theorem, each constraint $h_a(\vec{\theta})=0$ can in principle be used to eliminate a parameter from the problem, thus reducing the number of metrological degrees of freedom by one. Therefore, the number of effective independent parameters in the problem is the number $k$ of non-zero eigenvalues. If $k<d$ the QFIM is singular.

The natural question then is: what information is contained in the non-zero QFIM eigenvalues and their eigenvectors? 

\subsection{Re-parametrization}
Consider the general re-parametrization $\vec{\theta}\to \vec{\chi}$, where each new parameter $\chi_a = f_a(\vec{\theta})$ is a function $f_a$ of the original parameters $\vec{\theta}$. Here we consider a one-to-one bijective mapping with $d$ original parameters and $d$ new ones. We now employ the chain rule $\partial_{\theta_i} \hat{\rho} = (\vec{\nabla}_{\chi} \hat{\rho})\cdot (\partial_{\theta_i} \:\vec{\chi})$, where $\vec{\nabla}_{\chi}=(\partial_{\chi_1},\partial_{\chi_2},...)$ is the gradient operator in the transformed parameter space of $\vec{\chi}$, together with Eq.~\eqref{eq:QFIM_Entry} to write $\hat{\mathcal{I}}_{\theta_i,\theta_j} = \sum_{a,b}\hat{\mathcal{I}}_{\chi_a,\chi_b} \times(\partial_{\theta_i\:} \chi_a \:\partial_{\theta_j\:} \chi_b)$. Therefore we can write,
\begin{equation}\label{eq:Ireparam}
\hat{\mathcal{I}}(\vec{\theta}) = \hat{M}^T\: \hat{\mathcal{I}}(\vec{\chi}) \:\hat{M} \;,
\end{equation}
where $\hat{\mathcal{I}}(\vec{\chi})$ is the QFIM in the transformed parameters and $\hat{M}$ is the $d\times d$ Jacobian matrix with elements $\hat{M}_{b,j}=\partial_{\theta_j\:}\chi_b $. 
CRBs hold in terms of either set of parameters, ${\rm Cov}[\vec{\theta}] \ge (\mathcal{M}\hat{\mathcal{I}}(\vec{\theta}))^{-1}$ and ${\rm Cov}[\vec{\chi}] \ge (\mathcal{M}\hat{\mathcal{I}}(\vec{\chi}))^{-1}$.  
Eq.~\eqref{eq:Ireparam} is therefore consistent with the (linearized) error-propagation formula,
\begin{eqnarray}\label{eq:errorprop}
    {\rm Cov}[\vec{\chi}] = \hat{M} \:{\rm Cov}[\vec{\theta}]\: \hat{M}^{T} \;.
\end{eqnarray}

A special but important case arises when the re-parametrization $\vec{\theta}\to\vec{\chi}$ brings the QFIM $\hat{\mathcal{I}}(\vec{\chi})$ into diagonal form. In this case, the multi-parameter estimation problem reduces to a set of decoupled effective single-parameter estimation problems in transformed coordinates. 

Despite their apparent similarity, Eqs.~\eqref{eq:I_diagU} and \eqref{eq:Ireparam} express different concepts. The QFIM eigenvectors $\vec{u}_a$ in Eq.~\eqref{eq:I_diagU} constitute a basis satisfying the linear equation $\hat{\mathcal{I}}(\vec{\theta})\vec{u}_a=G_a\vec{u}_a$. But since the QFI depends non-linearly on the underlying parameters $\vec{\theta}$, the basis of eigenvectors contained in $\hat{U}$ does not directly correspond to the basis of \textit{parameters} $\vec{\chi}$ in terms of which the QFIM $\hat{\mathcal{I}}(\vec{\chi})$ is diagonal. Further analysis is required to extract this information. 
Indeed, we note that it may not always be possible to find a re-parametrization corresponding to the set of functions $\{f_a(\vec{\theta}) \}$ that yields a fully diagonal QFIM. That is, some multiparameter estimation problems may be irreducibly \textit{correlated}, with no representation in terms of fully independent effective parameters. 

Assuming that a diagonal QFIM $\hat{\mathcal{I}}(\vec{\chi})$ can be found, its elements can be compared with the QFIM eigenvalues by inserting Eq.~\eqref{eq:Ireparam} into Eq.~\eqref{eq:I_diagU} to obtain $\hat{\mathcal{D}}=(\hat{M}\hat{U})^{T}\:\hat{\mathcal{I}}(\vec{\chi})\:(\hat{M}\hat{U})$. Unless for some reason $\hat{M}=\hat{U}^{-1}$ we see that the QFIM eigenvalues are not just equal to elements of the transformed QFIM, but are linear combinations of these elements. We wish to emphasize that the Jacobian matrix $\hat{M}$ is not typically an orthogonal matrix and so it is indeed generally not the case that the QFIM eigenvalues correspond to effective single-parameter QFIs in some transformed basis. 

It is tempting to simply invert Eq.~\eqref{eq:Ireparam} to obtain $\hat{\mathcal{I}}(\vec{\chi}) = (\hat{M}^{-1})^T\: \hat{\mathcal{I}}(\vec{\theta}) \:\hat{M}^{-1}$ and then equate $\hat{\mathcal{I}}(\vec{\chi})=\mathcal{D}(\vec{\theta})$ in Eq.~\eqref{eq:I_diagU} to give the condition $\hat{M}(\vec{\theta})=\hat{U}(\vec{\theta})^{-1}$ and hence $\partial_{\theta_j}\chi_b = [\hat{U}(\vec{\theta})^{-1}]_{b,j}$. However, this is in general not possible because the Jacobian $\hat{M}$ is generally not orthogonal, and also may not be invertible. Thus, $\mathcal{D}(\vec{\theta})$ does not necessarily have the interpretation as a QFIM in some transformed basis of parameters $\vec{\chi}$. In the standard multi-parameter case, it is typically a difficult task to find the encoding $\{f_a(\vec{\theta}) \}$ to a set of transformed parameters in terms of which the QFIM is diagonal (if such a transformation exists).

\subsection{Re-parametrization for singular systems}

This has implications for extracting the effective parameters for a singular system from the QFIM.
For a singular QFIM, due to the over-parametrization we have $k<d$ estimable effective parameters $\vec{\Omega}$ corresponding to the non-zero eigenvalues of the QFIM, and $d-k>0$ metrologically ``dark'' parameters $\vec{\Lambda}$ corresponding to the zero eigenvalues. Together $\vec{\Omega}$ and $\vec{\Lambda}$ comprise the $d$ transformed parameters $\vec{\chi}$. 
This in turn implies that a transformation can be found in which the QFIM takes a block structure, with the only nonzero elements being those belonging to the parameters $\vec{\Omega}$. The block corresponding to the estimable effective parameters $\vec{\Omega}$ need not be diagonal.

A simplification arises when the general multi-parameter problem has only one nonzero eigenvalue and hence only one effective parameter $\Omega$ (that is, $k=1$ with $d-1$ zero eigenvalues).
In this case it can be shown that the single non-zero QFIM eigenvalue $G_1$ is given by,
\begin{eqnarray}\label{eq:qfim_eval}
G_1=\mathcal{I}_{\Omega}\times(\vec{\nabla}_{\theta}\Omega)^2  \;, 
\end{eqnarray}
with $\mathcal{I}_{\Omega}$ an effective single-parameter QFI. Furthermore, we find a condition on the corresponding QFIM eigenvector $\vec{u}_1$ satisfying $\hat{\mathcal{I}}(\vec{\theta})\vec{u}_1=G_1\vec{u}_1$, which can be expressed,
\begin{equation}\label{eq:evect_field}
    \vec{u}_1=\frac{\vec{\nabla}_{\theta}\Omega}{|\vec{\nabla}_{\theta}\Omega|} \;.
\end{equation}
For $k=1$ the QFIM eigenvector is therefore related to the \textit{gradient} of the corresponding effective parameter. The effective parameter can in this sense be viewed as a kind of metrological ``potential'' that sources the QFIM eigenvector ``field''. We give examples of the relation between the effective parameter and the corresponding QFIM eigenvector field in Appendix~\ref{app:triangle}. 

For this case with a single effective parameter, 
we have $d-1>0$ ``dark'' parameters $\vec{\Lambda}$ corresponding to the zero QFIM eigenvalues. These cannot be estimated. They relate to the metrological symmetries because changing a dark parameter $\Lambda_j(\vec{\theta})$ does not affect the measurement outcomes or our statistical model on which the estimators are based.

In principle, the QFIM can be diagonalized and then Eq.~\eqref{eq:evect_field} can be inverted to extract the parameter encoding for singular estimation problems; that is, the functional dependence $\Omega=f(\vec{\theta})$. However, in practice this is extremely difficult, because Eq.~\eqref{eq:evect_field} corresponds to a set of coupled nonlinear partial differential equations. An exact solution for $\Omega$ can only be obtained analytically from Eq.~\eqref{eq:evect_field} for the simplest of cases. We highlight examples where this can and cannot be done in the following sections. We also note that numerical solutions are  challenging due to numerical instabilities and branch-cut subtleties. 

The analysis on the level of the QFIM presented above indicates that information about the number of effective parameters can be straightforwardly obtained, but that constructing the estimator for these effective parameters is generally not feasible within this approach. Indeed, the analysis becomes considerably more complicated when there are $k>1$ effective parameters. For any practical metrological scheme, one is not only interested in the precision bound for some effective parameter, but also one must know the encoding that tells us what that parameter actually is in order to construct a viable estimator. This highlights the limitations of the standard Cram\'er-Rao formalism for singular multiparameter estimation problems.

Instead, we shall see that the Bayesian approach gives us direct access to the effective parameters in singular multiparameter estimation problems and allows us to construct the estimators, without needing to solve Eq.~\eqref{eq:evect_field}.  Furthermore metrological symmetries are vividly visualized, giving physical insight into the origin of metrological singularities and what information can really be learned from measurements on such a quantum system. 


\section{Physical origin of QFIM singularities}\label{sec:whensingular}
As discussed in the previous section, over-parametrization leads to singularities in the QFIM and the emergence of a \textit{reduced set} of metrological effective parameters. We now identify three main physical situations where this occurs.
\begin{enumerate}
    \item \textbf{Type 1 over-parametrization: state dependence on effective parameters.} In this type of over-parametrization, the quantum state of the probe depends on a reduced set of effective parameters and thus symmetries exist in the level of quantum state itself. Consequently, the data coming from measurements naturally reflect such symmetries and hence the QFIM becomes singular.

    \item \textbf{Type 2 over-parametrization: measurement dependence on effective parameters.} In this case, the optimal measurement bases, extracted from the SLDs, show metrological symmetries. Given the different dependences on the underlying parameters of the SLDs and the probe state, symmetries can in principle show up in measurements that are not present on the level of the state.   

    \item \textbf{Type 3 over-parametrization: hidden metrological symmetries.} In some instances, effective functional relationships between the original parameters are not discernable on the level of the quantum state of the probe or on the level of the SLDs. This can lead to highly non-trivial effective parameters. In this case, the quantum state and/or SLD operators can be parametrized in a genuinely independent manner by the original parameters, which nevertheless appear as indistinguishable signals in the measurement data and constructed statistical model. 
        
\end{enumerate}

In the following sections, we investigate these types of over-parametrization in detail. As shown in Sec.~\ref{sec:diag}, the frequentist approach is useful for revealing information about the \textit{number} of effective parameters, but it fails to uncover the exact functional form of the effective parameters in most practical settings. As we will see in the following sections, the Bayesian approach provides a natural framework for identifying the functional form of effective parameters and the metrological symmetries, without any \emph{a priori} knowledge of the precise encoding. The encodings naturally emerge in the posterior distribution, which then allows us to systematically estimate the values of these effective parameters through a re-parametrization of the problem.


\section{Type 1 over-parametrization: \\
state dependence on effective parameters}

Consider a quantum probe with the (pure or mixed) state $\hat{\varrho}$ which depends on the $d$ unknown parameters to be estimated $\vec{\theta}$ only through a smaller set of $k<d$ effective parameters $\vec{\Omega}$. We focus first on the simplest case of a \textit{single} effective parameter $\Omega=f(\vec{\theta})$ such that $\hat{\varrho}(\vec{\theta}) \equiv \hat{\varrho} \left( \Omega \right)$. This embodies a symmetry on the level of the states of the system, since different combinations of the underlying parameters $\vec{\theta}$ give the same physics provided that $\Omega$ is the same. The state is invariant to transformations $\vec{\theta}\to \vec{\theta}'$ that leave $\Omega$ unchanged. No measurement on the system can distinguish the parameters $\vec{\theta}$ from $\vec{\theta}'$ in this case. As an example consider a quantum state $\hat{\varrho}(\theta_1,\theta_2)=\hat{\varrho}(\theta_1/\theta_2)$, which implies that the effective parameter is $\Omega=\theta_1/\theta_2$. In this situation any transformation of $(\theta_1,\theta_2)\rightarrow (\eta\theta_1,\eta\theta_2)$ will not change the quantum state and thus gives the same measurement statistics.  It is straightforward to show that these types of symmetries directly imply a singular QFIM. Using the definition of the QFIM elements in Eq.~\eqref{eq:QFIM_Entry}, and the chain rule $\partial_{\theta_i}\hat{\varrho} \left( \Omega \right)=\partial_{\Omega}\hat{\varrho} \left( \Omega \right)\times \partial_{\theta_i}\Omega$, we find,
\begin{equation}
    \label{eq:effective_param_form}
    {\mathcal{I}}_{i,j}= \mathcal{I}_\Omega \times \left( \partial_{\theta_i} \Omega ~ \partial_{\theta_j} \Omega \right) \;,
\end{equation}
with $\mathcal{I}_\Omega = \sum_{k,l} 2\Re \langle e_k | \partial_{\Omega} \hat{\varrho}  | e_l \rangle^2 / (e_k + e_l)$ an \emph{effective} single-parameter estimation QFI for the effective parameter $\Omega$. An alternative derivation of Eq.~\eqref{eq:effective_param_form} can be obtained from Eq.~\eqref{eq:sld_explicit}. When $\hat{\varrho}(\Omega)$ depends only on $\Omega$, we may write $\hat{L}_i=\hat{L}_{\Omega}\times \partial_{\theta_i}\Omega$ in terms of the SLD operator for the effective parameter $\Omega$. Interestingly, we remark that this implies that SLD operators for different bare parameters $\theta_i$ and $\theta_j$ commute, and so the SLDs are certainly compatible. Although systems with compatible SLDs are often viewed as the ideal systems for multiparameter estimation, they are in fact singular when the state depends on a single effective parameter. 
We obtain Eq.~\eqref{eq:effective_param_form} from the second equality in Eq.~\eqref{eq:QFIM_sld} in this case, where now we can express the effective single-parameter QFI in the alternative form  $\mathcal{I}_{\Omega}=\text{Tr}\left( \hat{L}_\Omega \partial_{\Omega} \hat{\varrho}(\Omega)  \right)$. 

The factorized form of Eq.~\eqref{eq:effective_param_form} immediately implies that the QFIM is singular, $\text{det} [ \hat{\mathcal{I}} ]\! =\! 0$. To see this, note that Eq.~(\ref{eq:effective_param_form}) can be written as
\begin{eqnarray}\label{eq:factor}
\hat{\mathcal{I}}=\mathcal{I}_\Omega\times {\rm diag} (\vec{\nabla}_\theta \Omega)\:\hat{J}_d \:{\rm diag} (\vec{\nabla}_\theta \Omega) \;,
\end{eqnarray} 
where $J_d$ is the $d$-dimensional matrix of ones and $\vec{\nabla}_{\theta}=(\partial_{\theta_1},\partial_{\theta_2},...)$ is the gradient operator in parameter space. The determinant follows as ${\rm det}[\hat{\mathcal{I}}]=\mathcal{I}_\Omega^d \prod_i (\partial_{\theta_i}\Omega)^2\times {\rm det}[J_d]$. But since ${\rm det}[J_d]=0$, the QFIM is singular. This applies in any dimension $d$, whenever the state depends on a single effective parameter $\Omega$. A concrete example of this for the generic two-level system is provided in Appendix~\ref{app:2ls}.

As discussed above, a singular QFIM implies the existence of at least one zero eigenvalue. We can straightforwardly compute the eigenvalues of the QFIM by exploiting the representation Eq.~\eqref{eq:factor}. In this case it can be shown that the QFIM has a  \textit{single}  nonzero eigenvalue $G_1=\mathcal{I}_{\Omega}\times (\vec{\nabla}_{\theta}\Omega)^2$ and  $d-1$ zero eigenvalues. We note that the single non-zero eigenvalue of the QFIM is related to (but not \textit{equal} to) the QFI $\mathcal{I}_{\Omega}$ for the single effective parameter $\Omega$. The normalized QFIM eigenvector $\vec{u}_1$ corresponding to this non-zero eigenvalue is found from Eq.~\eqref{eq:effective_param_form} to be $\vec{u}_1=(\vec{\nabla}_{\theta} \Omega)/|\vec{\nabla}_{\theta} \Omega|$, in agreement with the general result of Eq.~\eqref{eq:evect_field}.

In the case where the state $\hat{\varrho}(\vec{\Omega})$ depends on $k$ effective parameters $\vec{\Omega}$, with $1\!<\!k\!<\!d$, we can generalize our expression for the QFIM. First, we note the chain rule  $\partial_i \hat{\varrho}(\vec{\Omega}) = (\vec{\nabla}_{\Omega} \hat{\varrho})\cdot (\partial_i \vec{\Omega})$ where $\vec{\nabla}_{\Omega}=(\partial_{\Omega_1},\partial_{\Omega_2},...)$ is the gradient operator in \textit{effective} parameter space. From Eq.~(\ref{eq:QFIM_Entry}) we then obtain,
\begin{equation}\label{eq:qfi_multiple_eff}
\begin{split}
\hat{\mathcal{I}}(\vec{\theta})&=\hat{T}^{\dagger} \hat{\mathcal{I}}(\vec{\Omega})\hat{T} \;,\\
&\equiv\sum_{a,b} \hat{\mathcal{I}}_{\Omega_a,\Omega_b} \times {\rm diag} (\vec{\nabla}_\theta \Omega_a)\:\hat{J}_d \:{\rm diag} (\vec{\nabla}_\theta \Omega_b) \;,
\end{split}
\end{equation}
where $\hat{\mathcal{I}}(\vec{\theta})$ is the $d\times d$ QFIM in terms of the bare parameters, $\hat{\mathcal{I}}(\vec{\Omega})$ is the $k\times k$ QFIM of effective parameters, and $\hat{T}$ is a $k\times d$ transformation matrix of derivatives, with elements $\hat{T}_{a,i}=\partial_i \Omega_a$. Analysis of these expressions shows that we have $k$ non-zero eigenvalues and $d-k$ zero eigenvalues, confirming our previous general results. Note that with multiple effective parameters $k>1$, the non-zero eigenvalues of the QFIM $\hat{\mathcal{I}}(\vec{\theta})$ are typically complicated linear combinations of elements of the effective QFIM $\hat{\mathcal{I}}(\vec{\Omega})$, and the effective parameter encoding is also obscure. For this, we benefit from the insights of Bayesian estimation.

In the following subsections, we demonstrate type 1 over-parametrization through instructive examples. We now focus on the two-parameter estimation scenario, $d=2$ with $\vec{\theta}=(\theta_1,\theta_2)$, where we have a single effective parameter $\Omega$. We present two contrasting cases, where in the first case the effective parameter $\Omega$ has an \textit{explicit}, known dependence on the underlying parameters $\theta_1$ and $\theta_2$, such as through a ratio, $\Omega=\theta_1/\theta_2$. In the second case, we consider an \textit{implicit} dependence where the encoding function $f$ for the effective parameter $\Omega=f(\theta_1,\theta_2)$ is \textit{a priori} unknown.


\subsection{Explicit parameter encoding}\label{sec:explicit}

We consider here an example in which the quantum state used for multiparameter estimation depends explicitly on only a single effective parameter.

We again take the $XY$ model Eq.~\eqref{eq:xy_general_ham},  and explore multi-parameter estimation of the field $h$ and coupling $\lambda$, this time assuming the anisotropy $\gamma$ is known. We present results for estimation using the ground state in the ring geometry for $N{=}4$ spins, but note that our results are representative of any $N$. For the ring geometry, elements of the QFIM are given by,
\begin{equation}
    \label{eq:xy_lam_h_qfim}
    \hat{\mathcal{I}}_{\lambda,h}^{\rm ring} = \sum_{k}\frac{16\sin^2(k)}{(\epsilon_k)^4}
    \begin{pmatrix}
   \; h^2 \gamma^2\;\; & -h \lambda \gamma^2 \;\\
   \; -h \lambda \gamma^2 \;\; & \gamma^2 \lambda^2 \; 
\end{pmatrix}\;,
\end{equation}
with $k =  \pi(2n+1)/N$ and $n = 0,1,2,\ldots,\left\lfloor\frac{N}{2}\right\rfloor - 1$ labelling the momentum space blocks, and  with dispersion $\epsilon_k = 2 \sqrt{(h + \lambda \cos{k} )^2 + \lambda^2\gamma^2 \sin^2{k}}$~~as before. In this case the QFIM is seen to be singular for all $N$, regardless of the values of the parameters $\vec{\theta}=(\lambda,h)$. 
Here the QFIM singularity is a direct result of Eq.~\eqref{eq:effective_param_form}, with an effective parameter $\Omega=h/\lambda$ explicitly defined. This choice can be immediately seen by rescaling the Hamiltonian as follows,
\begin{equation}
    \label{eq:xy_ham_rescaled_lam_h}
    \hat{H}' \equiv \frac{\hat{H}_{XY}^{\rm ring}}{\lambda} = \sum_{i = 1}^{N} \frac{1}{2} \left( \left[ 1 + \gamma \right] \hat{\sigma}_x^i \hat{\sigma}_x^{i+1} + \left[ 1 - \gamma \right]\hat{\sigma}_y^i \hat{\sigma}_y^{i+1} \right) + \sum_{i = 1}^{N} \frac{h}{\lambda} \hat{\sigma}_z^i\;.
\end{equation}
The rescaled Hamiltonian $\hat{H}'$ is dimensionless, with $\lambda$ being treated here as the unit of energy. Energy gaps between states of $\hat{H}_{XY}^{\rm ring}$ are typically of order $\mathcal{O}(\lambda/N)$ and so finite-temperature quantities involving a mixed state of the system are sensitive to the scale $\lambda$. On the other hand, pure states (such as the ground state) can depend on $\lambda$ only through the combination $\Omega=h/\lambda$, as seen directly from Eq.~\eqref{eq:xy_ham_rescaled_lam_h} and can be verified directly. It is then immediately obvious that $h$ and $\lambda$ cannot be independently estimated from pure states of the $XY$ model: no measurement using such a state can distinguish between $\vec{\theta}=(\lambda, h)$ and $(2\lambda, 2h)$ for example, because the \textit{ratio} of the parameters is the same in both cases, and the state only depends on the ratio.

With $\gamma$ known, Eq.~\eqref{eq:xy_ham_rescaled_lam_h} reduces to an effective single-parameter estimation problem for $\Omega=h/\lambda$. Even though the QFIM of the dual estimation problem is singular, we can deduce the existence of a well-defined effective single-parameter QFI for $\Omega$ via Eq.~\eqref{eq:effective_param_form}, given by $\mathcal{I}_{\Omega} = \sum_{k,l} 2\Re \langle e_k | \partial_\Omega \hat{\varrho} | e_l \rangle^2 /(e_k + e_l)$. For the ground state of the XY model on the ring geometry we obtain,
\begin{eqnarray}
\mathcal{I}_{\Omega}=\gamma^2\lambda^4\sum_{k}\frac{16\sin^2(k)}{(\epsilon_k)^4} \;. \label{eq:Iom_ratio}
\end{eqnarray}
The CRB holds for the effective parameter $\Omega$, which bounds the variance of estimators according to ${\rm Var}(\hat{\Omega})=1/(\mathcal{M}\mathcal{I}_{\Omega})$.

The direct result Eq.~\eqref{eq:Iom_ratio} is also consistent with the reparametrization Eq.~\eqref{eq:Ireparam} that takes 
$\hat{\mathcal{I}}(\vec{\theta})$  in Eq.~\eqref{eq:xy_lam_h_qfim} to $\hat{\mathcal{I}}(\vec{\chi})$, where $[\hat{\mathcal{I}}(\vec{\chi})]_{\Omega,\Omega}\equiv \mathcal{I}_{\Omega}$ and $\vec{\chi}=(\Omega,\Lambda)$ with $\Omega=h/\lambda$ the effective parameter of interest and $\Lambda=h\times \lambda$ the complementary `dark' parameter that encodes the metrological symmetry and cannot be estimated. 

The single non-zero eigenvalue of the QFIM is straightforwardly obtained from Eq.~\eqref{eq:xy_lam_h_qfim}, yielding  $G_1=\gamma^2(h^2+\lambda^2)\sum_{k}16 \epsilon_k^{-4} \sin^2(k)$. Comparison with Eq.~\eqref{eq:Iom_ratio} shows that  Eq.~\eqref{eq:qfim_eval} is satisfied (noting that $(\vec{\nabla}_{\theta}\Omega)^2=(h^2+\lambda^2)/\lambda^4$ in this case). The  corresponding QFIM normalized eigenvector is $\vec{u}_1=(-h,\lambda)/\sqrt{h^2+\lambda^2}$, which also satisfies Eq.~\eqref{eq:evect_field}.

We note that, in this simplest of cases where the effective parameter involves just a ratio of bare parameters, one can deduce $\Omega$ from $\vec{u}_1$ by analyzing Eq.~\eqref{eq:evect_field}. This amounts to solving the partial differential equation $(h^2+\lambda^2)[\partial_{\lambda}\Omega]^2=h^2([\partial_{\lambda}\Omega]^2+[\partial_{h}\Omega]^2)$ for $\Omega$. This yields $\Omega=f(h/\lambda)$ or $f(h\times \lambda)$. The latter leads to an inconsistency on the level of Eq.~\eqref{eq:qfim_eval}. Therefore we can conclude that the simplest effective parameter is indeed $\Omega=h/\lambda$. One might ask whether there is some function $\Omega=f(h/\lambda)$ such that the QFIM eigenvalue is precisely the effective QFI, $G_1=\mathcal{I}_{\Omega}$. However, it is easily proved that no such function exists. This again highlights that simply diagonalizing the QFIM does not (necessarily) yield a legitimate QFIM -- one has to find the proper re-parametrization.

\begin{figure}[t]
    \centering
    \includegraphics[width=1\linewidth]{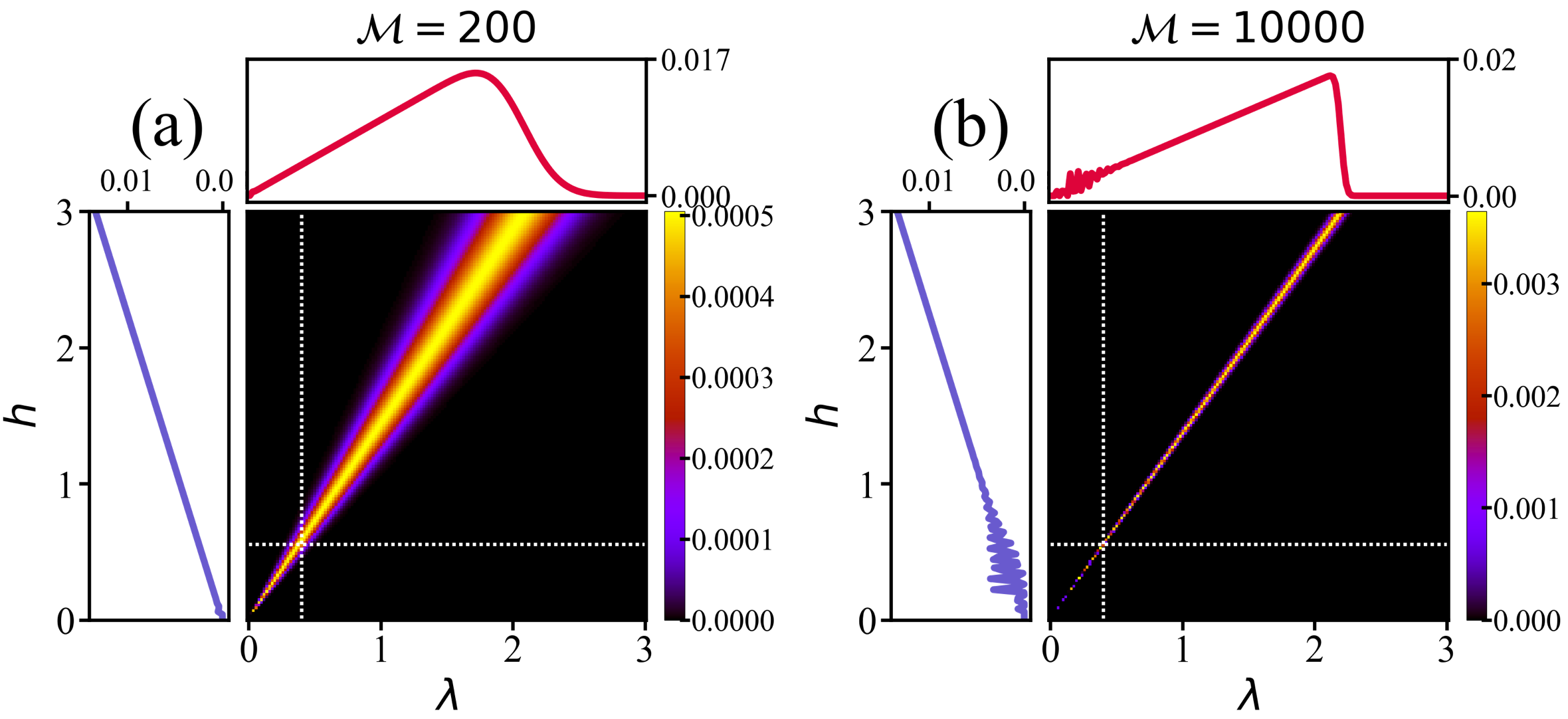}
    \caption{\textbf{Multinomial posterior distributions for Bayesian estimation of $\lambda$ and $h$ in the $XY$ model.} Results shown for estimation using the ground state of the $N=4$ system in the ring geometry, whose QFIM is singular. The true values of the parameters to be estimated (dotted white lines) is $\vec{\theta}^{\rm \:tr}\equiv (\lambda^{\rm tr}, h^{\rm tr})=(0.4, 0.55)$ and we set $\gamma=1$. Panels (a) and (b) compare the results after $\mathcal{M}=200$ and $10000$ measurements, respectively. Associated line plots show the individual (marginalized) posteriors for each parameter. }
    \label{fig:xy_ratio_post}
\end{figure}

One rather simple and vivid way to do this is to use Bayesian estimation. To showcase the power of the Bayesian approach for singular multi-parameter estimation problems, we now demonstrate that the above conclusions about the QFIM singularity, metrological symmetries, and effective parameter encoding can be drawn automatically, with no \textit{a priori} knowledge of the structure of the state. This is seen directly from the multinomial Bayesian posterior distribution. In Fig.~\ref{fig:xy_ratio_post} we present results for multi-parameter Bayesian estimation of $\lambda$ and $h$ in the $XY$ model, using the ground state of the $N\!=\!4$ system in the ring geometry, setting $\gamma\!=\!1$. We take the true values of the parameters to be estimated as $\vec{\theta}^{\rm \:tr}\equiv (\lambda^{\rm tr}, h^{\rm tr})=(0.4, 0.55)$, arbitrarily chosen. For the simulation, POVMs are again constructed from the eigenbasis of $\hat{\sigma}_z^{\text{tot}}$, which constitutes the optimal measurement for both parameters. Panels (a) and (b) compare the Bayesian posterior distributions after $\mathcal{M}=200$ and $10000$ measurements.

We immediately see that the QFIM singularity shows up in the Bayesian posterior distributions as a persistent line of high likelihood, whose width decreases with increasing $\mathcal{M}$. Note that the individual posteriors obtained after marginalizing (shown by the line plots above and to the left of the main panels) are distinctly non-Gaussian and have essentially no physical interpretation. The Bayesian averages (estimated values) of the parameters obtained from these do not converge to the true values of the parameters, shown by the dotted white lines, and are determined only by the sampling range.

The singular lines in Fig.~\ref{fig:xy_ratio_post} are described by the function $f(h,\lambda)\equiv h/\lambda = \Omega^{\rm tr}$ with the constant $\Omega^{\rm tr}=h^{\rm tr}/\lambda^{\rm tr}$ set by the true value of the parameters. We can therefore see directly from the Bayesian approach that there is a single effective parameter $\Omega=h/\lambda$ that characterizes the metrological problem. The true value $\Omega^{\rm tr}$ can be read off from the Bayesian posterior distributions, even though the individual values of $h^{\rm tr}$ and $\lambda^{\rm tr}$ remain ambiguous.

\subsection{Implicit parameter encoding} \label{sec:heisenberg}

We now consider a case where the state depends on a single effective parameter $\Omega$, but its underlying functional dependence on the bare parameters $\vec{\theta}$ is \textit{a priori} unknown. This implicit functional relationship will however be revealed in the Bayesian posterior.

For this purpose, we take as our example the Heisenberg spin-chain model. Here we will consider the dual estimation of a coupling constant (a Hamiltonian parameter) and the temperature (a non-Hamiltonian parameter). We therefore generalize now to mixed states at finite temperatures. Furthermore, we consider both the full state of the entire system as a resource for parameter estimation as well as the \textit{reduced} state, embodying partial accessibility to a restricted set of probe degrees of freedom. We shall see below that a number of nontrivial differences arise for the Heisenberg model, but that Bayesian estimation is able to provide detailed information about the fundamental structure of the metrological problem.

The Heisenberg model we consider consists of a one-dimensional chain of $N$ spins-$\tfrac{1}{2}$, with nearest-neighbour exchange couplings and open boundaries, \begin{equation}
    \label{eq:heisenberg_ham}
    \hat{H}_{\text{Heis}} = \sum_{i=1}^{N-1} J_i (\hat{\sigma}_x^{i} \hat{\sigma}_x^{i+1}+\hat{\sigma}_y^{i} \hat{\sigma}_y^{i+1}+\hat{\sigma}_z^{i} \hat{\sigma}_z^{i+1}) \equiv \sum_{i=1}^{N-1}J_i\: \vec{\boldsymbol{\sigma}}_i\cdot \vec{\boldsymbol{\sigma}}_{i+1}
\end{equation}
where $\boldsymbol{\sigma}_i=(\hat{\sigma}^i_x,\hat{\sigma}^i_y,\hat{\sigma}^i_z)$ is the Pauli vector. We take chains of even length $N$ and focus on the case where the central two spins are coupled by an exchange coupling $K$ and all other spins are coupled by $J$. Similar settings have been heavily studied for metrology, including the development of impurity-based probes~\cite{mihailescu2024critical,mitchell2012two,mitchell2011two}, non-equilibrium quantum sensors~\cite{rams2018limits}, optimal global thermometers~\cite{mok2021optimal}, multi-parameter quantum probes~\cite{bakmou2019quantum} and sequential measurement sensing schemes~\cite{OConnor_2024,montenegro2022sequential,yang2023extractable}. Here we assume the system is in thermal equilibrium and we focus on the joint estimation of $T$ and $K$ (with $J$ assumed to be known). For simplicity we will restrict to $N = 4$ spins, but remark that our results are rather generic for larger (finite, even) $N$. 

At finite temperatures, the full state of the system is described by a statistical mixture, $\hat{\varrho} = \sum_n e_n \ket{e_n}\bra{e_n}$ corresponding to the thermal Gibbs state. The probabilities $e_n$ are therefore the normalized Boltzmann weights $e_n=e^{-E_n/T}/Z$, with $Z=\sum_n e^{-E_n/T}$ the partition function and $E_n$ the energy of state $|e_n\rangle$ satisfying the Schr\"odinger equation $\hat{H}|e_n\rangle = E_n |e_n\rangle$. Information regarding the set of parameters is typically encoded in both the populations $e_n$ and eigenvectors $|e_n\rangle$ of the state $\hat{\varrho}$. 

We consider two settings: full accessibility to the complete spin chain and partial accessibility to only the two central sites. This allows us to examine how the level of access to the system impacts multi-parameter sensitivity. The reduced state of the central two spins is given by the partial trace, $\hat{\varrho}_{23}={\rm Tr}_{14}\{\hat{\varrho} \}$. For both the full and reduced state, the QFIM is given by Eq.~\eqref{eq:QFIM_Entry}. 

In the case where we use the full state of the entire system for metrology, the QFIM $\hat{\mathcal{I}}$ for $\vec{\theta}=(K, T)$ is non-singular at all finite temperatures $T$. In this case, the standard multi-parameter CRB scenario applies. By contrast, when the reduced state of the central two sites is used, the QFIM is singular. This can be traced back to the relatively small local Hilbert space of the `probe' spins, and the constraints imposed by $\text{SU}\left(2 \right)$ spin symmetry~\cite{mihailescu2024critical}. In fact, it is straightforwardly shown that the reduced density matrix (RDM) of the central sites $\hat{\varrho}_{23}$ is entirely specified by the thermal expectation value of the spin-spin correlation function between these sites, $\langle \hat{\vec{\bm{\sigma}}}_{\text{2}} \cdot \hat{\vec{\bm{\sigma}}}_{\text{3}} \rangle \equiv {\rm Tr}\{ \hat{\varrho}\:\hat{\vec{\bm{\sigma}}}_{\text{2}} \cdot \hat{\vec{\bm{\sigma}}}_{\text{3}} \}$, which controls the singlet-triplet fraction. 
Even though the correlation function $\langle \hat{\vec{\bm{\sigma}}}_{\text{2}} \cdot \hat{\vec{\bm{\sigma}}}_{\text{3}} \rangle =f(K, T)$ is some complicated (and \textit{a priori} unknown) function of the coupling $K$ and temperature $T$, on the level of the probe RDM it serves as an \textit{implicit} effective parameter $\Omega = \langle \hat{\vec{\bm{\sigma}}}_{\text{2}} \cdot \hat{\vec{\bm{\sigma}}}_{\text{3}} \rangle$ and in this case we can write $\hat{\varrho}_{23}\equiv\hat{\varrho}_{23}(\Omega)$. 
Appendix~\ref{app:heis_chain} shows a contour plot of $\Omega$ as a function of $T$ and $K$, and highlights the complicated implicit parameter relationships that can emerge in such systems. 

Since the reduced thermal state depends on a single effective parameter, Eq.~\eqref{eq:effective_param_form} holds and the QFIM is singular at all temperatures. This contrasts to the results of Sec.~\ref{sec:explicit} where the singularity arises only for pure states (see Sec.~\ref{sec:lift_sing} for a discussion on how the singularity is lifted at finite temperatures in the $XY$ model). 

\begin{figure}[t]
    \centering
    \includegraphics[width=1\linewidth]{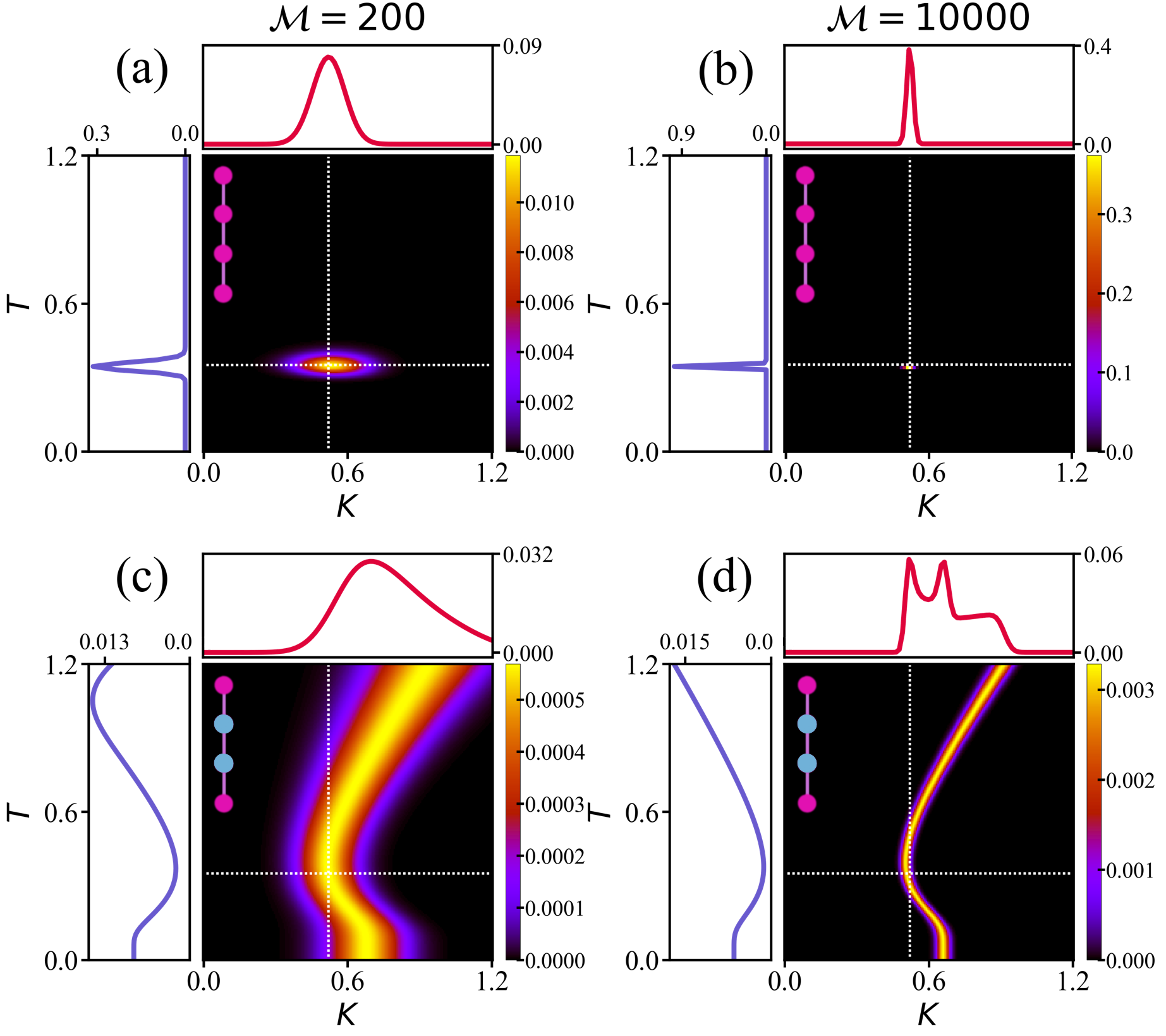}
\caption{\textbf{Multinomial posterior distributions for Bayesian estimation of $T$ and $K$ in the Heisenberg model.} For dual estimation of the temperature $T$ and spin coupling $K$, we use either the full thermal state of the entire system (top panels: a,b) or the reduced state of the central two spins (bottom panels: c,d). The QFIM is singular in the case of the reduced state. We compare $\mathcal{M}=200$ vs $10000$ measurements in the left and right panels, respectively. The true parameter values are $\vec{\theta}^{\rm \:tr}=\left( K^{\rm tr}, T^{\rm tr} \right) = \left( 0.52, 0.35 \right)$, shown as the white dotted line. Line plots show the individual (marginalized) distributions for each parameter. We use $N=4$ and $J=1$. }
    \label{fig:heisenberg_post}
\end{figure}

Another consequence of the SU(2) spin symmetry of the reduced state is that eigenstates of the RDM are the spin singlet and triplet states, independently of $K$ and $T$. Consequently, only the RDM populations have a parameter dependence, and the effective QFI simplifies to $\mathcal{I}_\Omega = \sum_n (\partial_\Omega p_n)^2/p_n$, where $p_n \equiv p_n(\Omega)$ are the reduced state populations.

For the Bayesian estimation strategy we fix $J = 1$ and take the true values of the parameters to be estimated as $\vec{\theta}^{\rm \:tr} \equiv (K^{\rm tr}, T^{\rm tr}) = (0.52, 0.35)$, arbitrarily chosen. The spin-spin correlation $\hat{\vec{\bm{\sigma}}}_{\text{2}} \cdot \hat{\vec{\bm{\sigma}}}_{\text{3}}$  is our observable of choice and POVMs are constructed from its eigenbasis. For estimation of $T$ and $K$ using the thermal state of the entire system, this choice of observable constitutes a sub-optimal measurement, i.e.~the measurement operator does not commute with individual parameter SLDs. The Fisher information matrix associated with this particular observable is however invertible and therefore multi-parameter Bayesian estimation is well-defined and should converge to the true parameter values, albeit not with optimal precision scaling. When we instead use the reduced thermal state of the central two `probe' spins for estimation, the spin-spin correlation $\hat{\vec{\bm{\sigma}}}_{\text{2}} \cdot \hat{\vec{\bm{\sigma}}}_{\text{3}}$  constitutes an optimal measurement for both parameters. However in this case, the corresponding QFIM is singular, as noted above. Therefore we do not expect the Bayesian posterior distribution to converge to a unique point in parameter space. In both cases, we take our initial prior as a uniform distribution over the range  $0 \leq \theta \leq 2$ for $\theta = T$ or $K$.

In the main panels of Fig.~\ref{fig:heisenberg_post} we show the resulting multinomial posterior distribution obtained after  $\mathcal{M}\!=\!200$ and $\mathcal{M} = 10000$ measurements in panels (a,c) and (b,d) respectively. The top row panels (a,b) are for estimation with the (non-singular) full state; bottom row panels (c,d) are for the (singular) reduced probe state case. White dotted lines show the true values of the parameters. Associated line plots show the individual (marginalized) posterior distributions for each parameter obtained from Eq.~\eqref{eq:individual_posterior}. 

Our simulation results in Fig.~\ref{fig:heisenberg_post} show very clearly that for the non-singular (full state) case (panels a,b) the Bayesian posterior distribution converges to a single unique point corresponding to the true parameter values. However, in the singular case (reduced state) in panels (c,d) the Bayesian posterior converges instead to a persistent \textit{line} of high likelihood. In this example of the Heisenberg chain model, the singular line has a nontrivial functional dependence on the parameters to be estimated.  The marginalized distributions in this case are non-Gaussian and lack a clear physical interpretation; the corresponding Bayesian average does not converge to the true values of the parameters as the number of measurements $\mathcal{M}$ is increased. Importantly though, the singular line of high likelihood does pass through the point corresponding to the true values of the parameters.

We see directly from the Bayesian posterior distributions for the case of estimation with the reduced state, Fig.~\ref{fig:heisenberg_post}(c,d), that measurements on the probe cannot distinguish between different combinations of the parameters $K$ and $T$ that lie on the persistent line. This indicates a metrological symmetry, where measurements are invariant to changes in $K$ and $T$ provided $\Omega = f(K,T)$ remains constant. The contour lines of constant $\Omega$ are discussed in Appendix~\ref{app:heis_chain} and are shown in Fig.~\ref{fig:Contour_Corr}(b).  Because the persistent line passes through the true values of the parameters, the line is defined by the equation $\Omega \equiv f(K,T)=\Omega^{\rm tr}$ with the constant $\Omega^{\rm tr}=f(K^{\rm tr}, T^{\rm tr})$ fixed by the true values. This plays the role of a constraint between the bare parameters $K$ and $T$. Even though the temperature $T$ and coupling constant $K$ are manifestly independent (in the sense that we can fix $T$ say, and still vary $K$ fully and freely), they become constrained and implicitly related on the level of the metrological statistical model. This leads to a redundancy in the parameter encoding and hence over-parametrization. At heart, this causes the QFIM singularity for this problem. Since the Bayesian posterior converges to the correct contour line, $\Omega^{\rm tr}$ can be estimated by the Bayesian strategy.


\section{Type 2 over-parametrization:\\
measurement dependence on effective parameters}

In principle, the optimal measurements corresponding to the SLD operators $\hat{L}_i$ may depend only on the $d$ parameters $\vec{\theta}$ through a reduced set of $k$ effective parameters $\vec{\Omega}$. This can happen even when the underlying state $\hat{\varrho}$ has no such reduced dependence (that is, the parameters $\vec{\theta}$ are \textit{independent} on the level of the state but not on the level of the measurements). 

For simplicity we again consider the two-parameter estimation case. Suppose that the SLD operators $\hat{L}_i$ depend only on the parameters $\theta_1$ and $\theta_2$ through the single effective parameter $\Omega$. Thus $\hat{L}_i(\theta_1,\theta_2)\equiv \hat{L}_i(\Omega)$ for all $i$, and the measurement data cannot distinguish between systems with parameters $(\theta_1,\theta_2)$ and $(\theta'_1,\theta'_2)$ provided they both correspond to the same effective parameter $\Omega$. In this case one can straightforwardly show that the QFIM is singular. This time we use Eq.~\eqref{eq:QFIM_sld} and the chain rule $\partial_{\theta_i}\hat{L}_j \left( \Omega \right)=\partial_{\Omega}\hat{L}_j \left( \Omega \right)\times \partial_{\theta_i}\Omega$ to write,
\begin{equation}
    \label{eq:effective_sld}
    \mathcal{I}_{i,j} = \mathcal{L}_j^\Omega \times \partial_{\theta_i} \Omega
\end{equation}
where $\mathcal{L}_j^\Omega = - \text{Tr}\{ \hat{\varrho}(\vec{\theta}) \partial_\Omega \hat{L}_j(\Omega) \}$. From this we obtain $\text{det} [ \hat{\mathcal{I}} ] = 0$. We emphasize that this parametric redundancy on the level of \textit{measurements} can happen even when the state $\hat{\varrho}$ does not have any such simple dependence. This can be seen from the fact that the parametric dependence of the SLDs is different from that of the state, see Eq.~\eqref{eq:sld_explicit}.
Although a distinct mathematical possibility, finding an explicit example of this type of QFIM singularity is an open question.


\begin{figure}[t]
    \centering
    \includegraphics[width=1\linewidth]{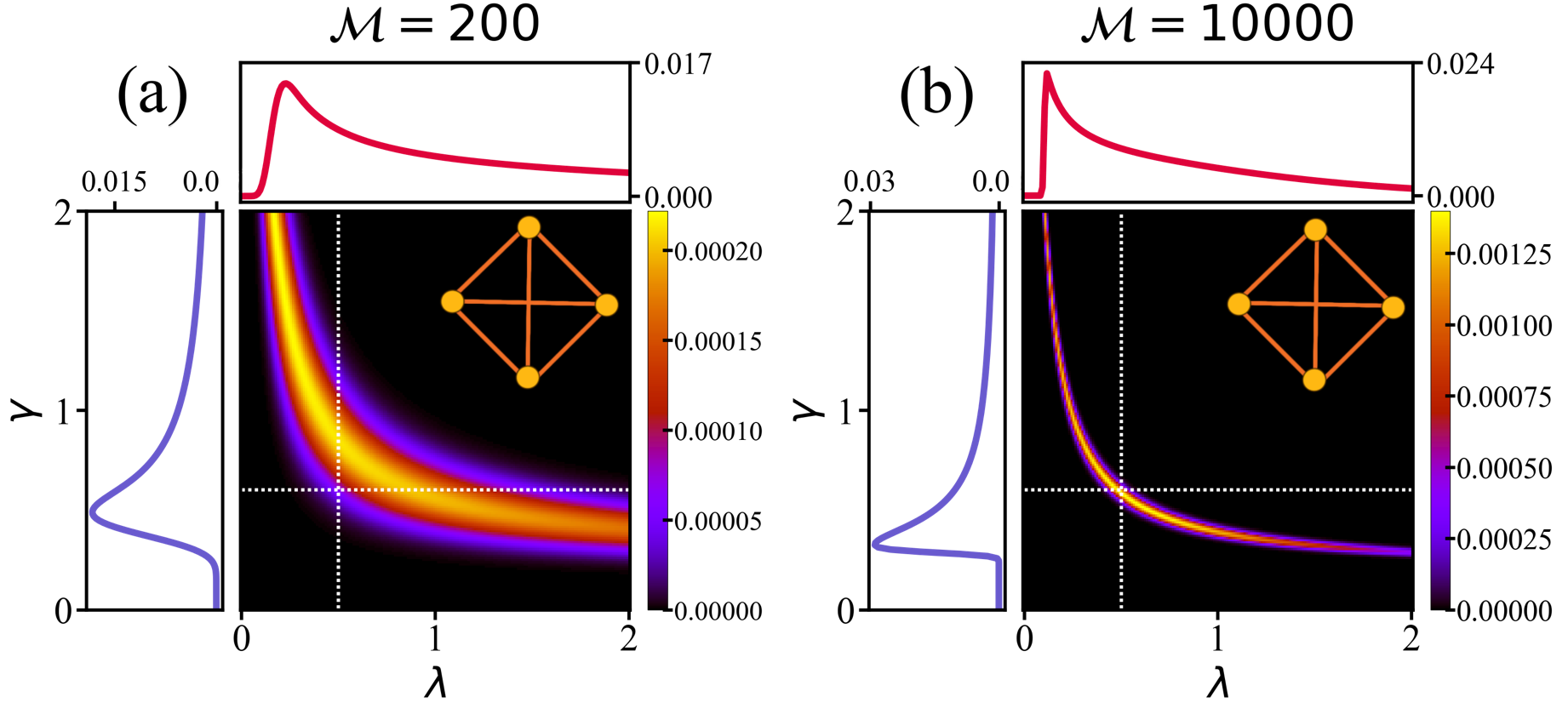}
    \caption{ \textbf{Multinomial posterior distributions for Bayesian estimation of $\lambda$ and $\gamma$ from the ground state of the $XY$ model with all-to-all coupling.} Shown for $\mathcal{M} = 200$  and $10000$  measurements in panels (a) and (b) respectively. White dotted lines show the true parameter values at $\left( \lambda^{\rm tr}, \gamma^{\rm tr} \right) = \left( 0.5,0.6\right)$. Line-plots show the corresponding marginalized univariate distributions for each parameter. Results shown for $N=4$ spins and known constant field $h=1$.}
    \label{fig:xy_banana}
\end{figure}


\section{Type 3 over-parametrization:\\
hidden metrological symmetries}\label{sec:hidden}

We note that both Eqs.~\eqref{eq:effective_param_form} and \eqref{eq:effective_sld} yield QFIM singularities by virtue of a factorization of the form $\hat{\mathcal{I}}\!=\!{\rm diag}(\vec{A})\:\hat{J}_d\:{\rm diag}(\vec{B})$, where $\vec{A}$ and $\vec{B}$ are $d$-dimensional vectors and $\hat{J}_d$ is the $d$-dimensional matrix of ones. Then ${\rm det}[\hat{\mathcal{I}}]=0$ since ${\rm det}[\hat{J}_d]=0$. 
These cases demonstrate the intimate connection between QFIM singularities and the existence of effective parameters or metrological symmetries. However,  there may be other situations where QFIM singularities arise, but no such simple factorization is possible. 

To demonstrate this third scenario, we again consider the $XY$ model of $N$ spins, Eq.~\eqref{eq:xy_general_ham}, but this time with uniform all-to-all interactions, $\lambda_{ij} \equiv \lambda / N$ and $\gamma_{ij} \equiv \gamma / N$ for all $i,j$. In this case we can recast Eq.~\eqref{eq:xy_general_ham} in terms of collective spin operators $\hat{S}_{\alpha} \equiv \sum_i \hat{\sigma}_{\alpha}^i / 2$. The Hamiltonian then reads,
\begin{align}
    \label{eq:xy_A2A_limit}
    \hat{H}_{XY}^{A2A} &= \lambda \left[ \left( 1 + \gamma\right) \hat{S}_x^2 
    + \left( 1 - \gamma \right) \hat{S}_y^2 \right] + 2 h \hat{S}_z\;. 
\end{align}
The total spin $S$ is conserved by $\hat{H}_{XY}^{A2A}$ and the ground state lives in the extremal spin sector $S=N/2$ throughout the parameter regimes considered for this model ($h=1$ and $\lambda,\gamma>0$). 

In contrast to the ring-geometry, the QFIM for joint estimation of $\lambda$ and $\gamma$ is singular for the all-to-all geometry, with ${\rm det}\left[\hat{\mathcal{I}}_{\gamma,\lambda}^{\rm A2A}\right] =0$. In this case, there is no parametric redundancy or single effective parameter that one can deduce directly from the structure of the ground state or the SLD operators for $N>3$. Yet the singular QFIM implies a single zero QFIM eigenvalue and therefore a single ``hidden'' metrological effective parameter $\Omega$.

As shown in Appendix~\ref{app:xyn3}, simple expressions for the QFIM can be obtained $N=3$ spins, and the explicit form of the effective parameter extracted analytically. For $N \ge 4$ however, there is no simple analytic form for the QFIM, and we were not able to invert Eq.~\eqref{eq:evect_field} to find $\Omega$ this way.

Instead, to uncover the form of the effective parameter and the nature of the metrological symmetry for $N=4$, we turn to Bayesian estimation. Convergence to the true values of $\lambda$ and $\gamma$ is no longer guaranteed (or expected) in this situation, but we still anticipate that useful metrological information will be extractable from the  Bayesian posterior distribution.  We note that, similar to the previous cases, the SLDs also commute in this case, implying that we we have a compatible but singular multi-parameter estimation problem. This arises even though the state depends separately on both parameters $\lambda$ and $\gamma$.

Because the QFIM is singular, the Bayesian posterior distribution is characterized by persistent regions of high likelihood, rather than convergence to a unique point corresponding to the true values of the parameters $\lambda^{\rm tr}$ and $\gamma^{\rm tr}$. Importantly, we observe that these regions of high likelihood become narrower with successive updates of the prior and converge to a specific \textit{line} in the space of bare parameters $\lambda$ and $\gamma$. This is shown in  Fig.~\ref{fig:xy_banana}, in which we compare the results of $\mathcal{M}=200$ and $10000$ measurements in panels (a) and (b).  We interpret such persistent lines in the posterior distribution as meaning that measurements cannot distinguish between different combinations of the parameters $\lambda$ and $\gamma$ along these lines -- an emergent metrological symmetry.

Importantly, the ground state used for estimation here \textit{does not} reflect this metrological symmetry (we have verified this directly by numerical calculation). Therefore the over-parametrization in this case is indeed of type 3.

The individual posteriors for $\lambda$ and $\gamma$, obtained after marginalizing the full multinomial distribution, are clearly non-Gaussian -- see associated lineplots in Fig.~\ref{fig:xy_banana}. Although Bayesian estimators for the parameters can still be constructed according to Eq.~\eqref{eq:bayesian_average}, inferences drawn from them are meaningless because of the ambiguity along the persistent lines of likelihood. Naive application of the Bayesian estimation strategy therefore does not yield correct estimates for the unknown parameters when the QFIM is singular.

Importantly however, note that the true values of the parameters $(\lambda^{\rm tr},\gamma^{\rm tr})$ do lie on the singular line in the Bayesian posterior. Therefore, although from the data we cannot determine uniquely the true values of the individual parameters, Bayesian estimation still provides useful new information. We learn that the true values of the parameters $\lambda^{\rm tr}$ and $\gamma^{\rm tr}$ to be estimated lie on a specific line $f(\lambda,\gamma)=\Omega^{\rm tr}$, where the constant $\Omega^{\rm tr}=f(\lambda^{\rm tr},\gamma^{\rm tr})$ is itself determined by the true parameters. The functional form $f(\lambda,\gamma)$, which embodies a kind of metrological constraint between the individual parameters $\lambda$ and $\gamma$, need not be known \textit{a priori}, but emerges automatically in the Bayesian posterior distribution after repeated updates. We emphasize that this functional relationship can be nontrivial and might not be easily guessed from analysis of the bare model or SLD operators, as in the present case. 

In the specific example of the all-to-all connected XY model, we find that the singular lines in the Bayesian posterior are quite accurately (although not precisely) described by the function $f(\lambda,\gamma)\equiv \lambda\times\gamma=\Omega^{\rm tr}$, with the constant $\Omega^{\rm tr} =\lambda^{\rm tr}\times\gamma^{\rm tr}$, and therefore describe hyperbolic contours in the $(\lambda,\gamma)$ plane. The Bayesian posterior allows us to read off the functional dependence of the effective parameter $\Omega \simeq \lambda\times\gamma$. Indeed, as shown in the next section, the true value $\Omega^{\rm tr}$ can be estimated with a precision that is determined by an effective CRB.

\begin{figure*}[t]
\centering
    \includegraphics[width=1\textwidth]{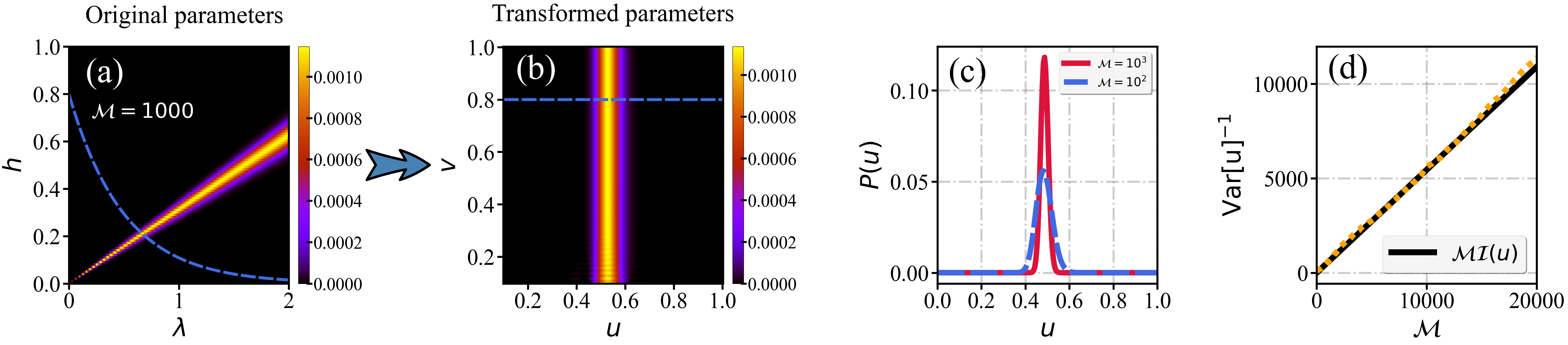}
 
    \includegraphics[width=1\textwidth]{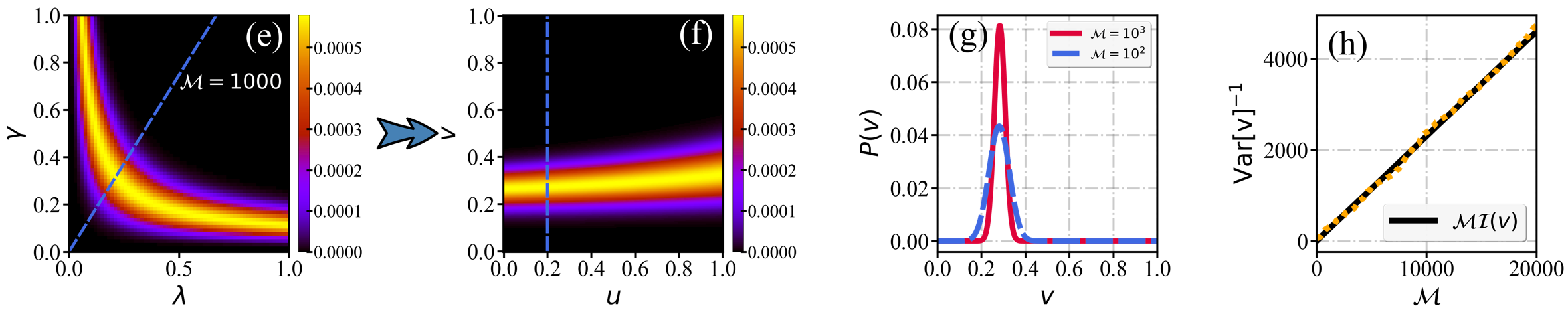}
\caption{\textbf{Recovery of effective CRB for singular multi-parameter estimation problems using co-ordinate transformations.} Results presented for the $XY$ model ground state with $N{=}4$ spins, showing dual estimation of $\vec{\theta}=(\lambda,h)$ for the ring geometry in the top row panels (a-d), and dual estimation of $\vec{\theta}=(\lambda,\gamma)$ for the all-to-all geometry in the bottom row panels (e-h). The QFIM in both cases is singular. The true values of the parameters are taken to be $\left( \lambda^{\rm tr}, h^{\rm tr} \right) = \left( 0.7,0.22\right)$ and $\gamma=1$ for panels (a-d); and $\left( \lambda^{\rm tr}, \gamma^{\rm tr} \right) = \left( 0.3,0.2\right)$ and $h=1$ for panels (e-h). The multinomial Bayesian posteriors in terms of the original parameters are presented in panels (a) and (e) for the two cases, both corresponding to $\mathcal{M}=1000$ measurements. The blue-dashed lines show line cuts through the persistent likelihood regions which have approximately Gaussian profiles. Panels (b) and (f) correspond to the same distributions plotted in terms of the transformed (hyperbolic) effective parameters $u$ and $v$ (see text). Blue-dashed lines are the same line cuts as in (a) and (e). Panels (c,g) show the corresponding marginalized univariate distributions, which are Gaussian, for $\mathcal{M}=100$ (blue lines) and $\mathcal{M}=1000$ measurements (red lines), and show a precision gain in the estimation of the effective parameter as more data are collected. The inverse variance of these distributions is extracted and plotted vs $\mathcal{M}$ in panels $(d,h)$ as the solid black lines, comparing with the effective single-parameter CRB for the transformed parameters (orange-dotted line).}
\label{fig:hyperbolic_trans}
\end{figure*}

\section{Effective Cram\'{e}r-Rao bound for\\singular estimation problems}\label{sec:effecitve_crb}

Independently of the type of QFIM singularity or the particular model setting, we find that the Bayesian posterior distribution for two-parameter estimation converges in the limit of a large number of measurements to a \textit{line} in parameter space, and not a unique \textit{point} as for non-singular estimation problems \footnote{For a $d$-dimensional estimation problem with $k$ non-zero QFIM eigenvalues (and therefore $k$ effective parameters), the Bayesian posterior converges to a $d-k$ dimensional submanifold.}. 
The functional form of this line encodes a metrological \textit{symmetry}: any system along this line yields asymptotically the same set of measurement outcomes and therefore the same statistical model. Different systems along this line cannot be distinguished. On the other hand, the lines perpendicular to this singular line encode the metrological \textit{effective parameter}. Measurements on the system give different results for different values of this parameter, which can therefore be estimated via the Bayesian strategy.

In this section, we develop a formalism to systematically identify the metrological symmetries and effective parameters from Bayesian analysis. By performing a re-parametrization to new coordinates, we show that an effective CRB holds for the single effective parameter, and demonstrate with specific examples that the true value of this effective parameter can indeed be reliably estimated -- even though the original \textit{multiparameter} estimation problem is singular. 
We emphasize that there is in general no straightforward strategy for determining the effective parameter encoding or metrological symmetries just from consideration of the model or its QFIM.

Understanding QFIM singularities as resulting from over-parametrization, the multinomial Bayesian posterior should be interpreted as a \emph{univariate} Gaussian distribution when the problem is characterized by a single effective parameter $\Omega$. In the limit of a large number of measurements $\mathcal{M}$, the Bayesian posterior in such cases can be approximated by,
\begin{equation}
    \label{eq:gaus_univ}
    P(\Omega \mid \{ n_k \}) \approx \sqrt{\frac{\mathcal{M} \mathcal{I}(\Omega^{\rm tr})}{2 \pi}} \exp{-\frac{ \mathcal{M} \mathcal{I}(\Omega - \Omega^{\rm tr})^2}{2}}\;.
\end{equation}
Here the effective parameter $\Omega = f(\theta_1,\theta_2)$ is some function of the  original parameters $\theta_1$ and $\theta_2$. 
The posterior distribution in the space of these original parameters allows us to ascertain the mapping between them and the effective parameter $\Omega$ directly and without any \emph{a priori} knowledge.

The individual (marginalized) posteriors for each bare parameter are not asymptotically normal -- see e.g.~Figs.~\ref{fig:xy_ratio_post} and \ref{fig:xy_banana}. Fundamentally this is because several combinations of $\theta_1$ and $\theta_2$ yield the same $\Omega$. However, as we shall demonstrate explicitly, the posterior for the \textit{effective} parameter $P(\Omega | \{ n_k \})$ does tend to a Gaussian, as per Eq.~\eqref{eq:gaus_univ}. Note that the persistent lines of high likelihood in the Bayesian posterior are \textit{contour} lines along which the effective parameter $\Omega$ is constant. If we had chosen different values of the bare parameters, we would get a different contour line, this time passing through the point corresponding to the new set of true parameter values. Measurements on the system along any such line must yield the same results. Therefore the underlying statistical model is invariant to changes in an orthogonal, independent parameter $\Lambda$ describing the position along those lines. We have a dependence on $\Omega$ but not $\Lambda$. We argue that taking a linecut through $\Omega$ while keeping $\Lambda$ constant should yield a Gaussian profile. To obtain a properly defined marginalized distribution, one must integrate all such orthogonal cuts along the singular line in question. In terms of the original coordinates, this procedure would typically be very cumbersome.

This motivates a generalized coordinate transformation $(\theta_1,\theta_2)\to (u,v)$ such that $u=\Omega$ is defined as being the effective parameter on which the underlying statistical model depends. The other parameter $v=\Lambda$ is independent of $u$. Transforming the Bayesian posterior distribution to $(u,v)$ coordinates should give perfect translational invariance along $v$. Marginalizing over $v$ gives the posterior $P(u)$ which should be Gaussian.

We now illustrate these concepts with simple examples from the $XY$ model, Eq.~\eqref{eq:xy_general_ham}. Our results are summarized in Fig.~\ref{fig:hyperbolic_trans}. In panel (a) we reproduce the Bayesian posterior distribution after $\mathcal{M}=1000$ measurements for dual estimation of $h$ and $\lambda$ in the $XY$ model for $N=4$ assuming a ring geometry; and in panel (e) we show the analogous plot for estimation of $\lambda$ and $\gamma$ for the all-to-all geometry. In both settings a singular QFIM is captured by the Bayesian posterior through the persistent lines of high likelihood. For panel (a) we deduce the existence of an effective parameter $\Omega=h/\lambda$; in panel (e) we assign $\Omega\simeq \lambda\times\gamma$. The equation $\Omega=\Omega^{\rm tr}$ defines the singular lines. In both cases the dependence of the effective parameter $\Omega=f(\theta_1,\theta_2)$ on the bare parameters $\theta_1$ and $\theta_2$ suggests a standard hyperbolic coordinate transformation,
\begin{equation}
    \label{eq:hb_trans}
    u = \ln{\sqrt{\frac{\theta_1}{\theta_2}}}, \qquad    v = \sqrt{ \theta_1 \theta_2}
\end{equation}

For the top row panels in Fig.~\ref{fig:hyperbolic_trans} we choose $u=\ln\sqrt{h/\lambda}$ and $v=\sqrt{h\times\lambda}$. Measurements on the system are sensitive to variations in $u$ but not $v$. This is in fact an exact statement: the ground state used for estimation here depends explicitly only on $u$ and not $v$. Panel (b) shows the Bayesian posterior as a function of these transformed coordinates. The hyperbolic linecut at constant $v$ is shown as the blue dashed line in panel (a). This same cut is a straight line in panel (b). As expected, we see perfect translational invariance along $v$ in panel (b). Marginalizing over $v$ yields $P(u)$ which we plot vs $u$ in panel (c), for two different values of $\mathcal{M}$. Two immediate observations are that the univariate distributions are indeed Gaussian, and that their width decreases with $\mathcal{M}$. We extract the precision ${\rm Var}[\hat{u}]^{-1}$ from $P(u)$ and plot this vs $\mathcal{M}$ in panel (d). This shows perfect linear scaling with the number of measurements, consistent with an effective CRB for the effective parameter $u$. To confirm this, we compute the single-parameter QFI $\mathcal{I}(u)$ for the $XY$ model, and plot $\mathcal{M}\mathcal{I}(u)$ as the orange-dotted line in Fig.~\ref{fig:hyperbolic_trans}(d). The exact agreement at large $\mathcal{M}$ demonstrates that we do indeed have an effective CRB, controlled by an effective single-parameter QFI, even in this singular multi-parameter estimation problem. 

We further illustrate these ideas in the bottom row panels (e-h) of Fig.~\ref{fig:hyperbolic_trans}, for the (singular) case of multi-parameter estimation of $\lambda$ and $\gamma$. In this case, we choose $u=\ln\sqrt{\lambda/\gamma}$ and $v=\sqrt{\lambda\times\gamma}$. Our ansatz is that the underlying statistical model this time depends only on $v$ and not on $u$. However, we note that this is deduced approximately from the structure of the singular line in panel (e) -- the state and SLD operators appear to still depend on both $u$ and $v$. Performing the hyperbolic transformation $(\lambda,\gamma)\to (u,v)$ and plotting the resulting Bayesian posterior in panel (f), we see that the distribution is indeed close to Gaussian along $v$ and close to invariant along $u$. However, this is seen to be an approximation, and there is some small dependence on $u$ evident from the plot. The linecuts at constant $u$ shown as the dashed lines in panels (e,f) yield Gaussian profiles to a good approximation. Marginalizing over $u$ gives the posterior $P(v)$ plotted in panel (g). Again, to a good approximation, these are Gaussian. Their width is seen to decrease with increasing $\mathcal{M}$ as before. Plotting the precision  ${\rm Var}[\hat{v}]^{-1}$ vs $\mathcal{M}$ in panel (h) again shows CRB scaling, set by an effective single-parameter QFI $\mathcal{I}(v)$, which we compute directly from the bare model (orange-dotted line). Although the true effective parameter appears not to be precisely $v=\sqrt{\lambda\times\gamma}$, corrections to this form are rather small.

Our results show that when the effective parameter is known precisely, the transformation can be performed  to give an asymptotically exact univariate Gaussian distribution for the Bayesian posterior at large $\mathcal{M}$, that follows an effective CRB. For this effective single-parameter problem, the BVM theorem holds. Indeed, even when the effective parameter is not known exactly, the structure of the Bayesian posterior still allows for an approximate mapping, based on an ansatz for the functional dependence inferred from measurement data.

We conclude from this that the fact of a singular QFIM does not preclude Bayesian parameter estimation -- rich information can still be learned. Furthermore, data can still be collected in the usual way, despite persistent lines of Bayesian likelihood signalling estimation ambiguities: one can simply post-process the data via appropriate parameter transformations as per Fig.~\ref{fig:hyperbolic_trans}. However, in such cases, only information about the effective parameters, and not the bare parameters, can be unambiguously determined. 

Finally, we remark that the introduction of non-linear coordinate transformations in this way for dealing with over-parametrization may have implications in the field of distributed sensing. In such scenarios, multiple users, each with access to only a subsystem of the probe, are involved in estimation of an effective global parameter of interest. This global effective parameter is typically a \textit{linear} combination of local parameters that affects the quantum state of the probe locally \cite{Rubio_2020,PhysRevResearch.3.033011,PhysRevLett.120.080501,proctor2017networkedquantumsensing}. As with the transformations illustrated in this section, it may be possible to generalize this to \textit{non-linear} parameter encodings.


\subsection{Effective CRB vs pseudoinverse method}
One approach that has previously been used to deal with singular multiparameter estimation problems is to apply the Moore-Penrose pseudoinverse to the FIM or QFIM~\cite{dtamin,stoica2001parameter}. Although the true matrix inverse $\hat{\mathcal{I}}(\vec{\theta})^{-1}$ of a singular QFIM does not exist, its pseudoinverse $\hat{\mathcal{I}}(\vec{\theta})^{+}$ always exists and is well-defined. Interpreting the QFIM pseudoinverse as a lower bound ${\rm Cov}(\vec{\theta})\ge (\mathcal{M}\hat{\mathcal{I}}(\vec{\theta}))^{+}$ on the covariance matrix of parameters to be estimated akin to the CRB is however known to be overly optimistic~\cite{xavier2004riemannian}. Yet the simplicity of the approach is appealing.

In light of the physical insights gained from the Bayesian analysis presented above, one might ask how meaningful the QFIM pseudoinverse is in this context. For example, from Figs.~\ref{fig:hyperbolic_trans}(a,e) we learn that, for the considered models, complete uncertainty in one parameter $\theta_1$ implies that the other parameter $\theta_2$ could take any value. The estimator $\hat{\theta}_2$ for that parameter will have infinite variance because for every possible value of $\theta_2$ there is some corresponding value of $\theta_1$ that gives the same effective $\Omega$. No measurement can statistically distinguish between systems with the same $\Omega$. Within a finite sampling window, the expected value and variance of the estimators is of course finite, but these values are only controlled by the details of the sampling window and do not give a proper estimate of the true values.  Therefore, the finite elements of the QFIM pseudoinverse appear to have very little physical significance in this case.

On the other hand, Figs.~\ref{fig:hyperbolic_trans}(d,h) show that a well-defined QFI and CRB do hold for the single effective parameter $\Omega$. Unfortunately, the QFIM pseudoinverse gives no hint about what this effective parameter is, and how it is related to the original parameters of the problem $\vec{\theta}$. Interestingly, however, the pseudoinverse does capture the correct precision bound for the effective parameter, as now shown.

Taking the case of Figs.~\ref{fig:hyperbolic_trans}(a-d) as a concrete example in which the effective parameter is identified as $\Omega=h/\lambda$, the CRB yields ${\rm Var}(\hat{\Omega}) \ge (\mathcal{M}\mathcal{I}_{\Omega})^{-1}$ which is well-defined and finite. Here  $\mathcal{I}_{\Omega})$ is given by Eq.~\eqref{eq:Iom_ratio}. Meanwhile, from the error propagation formula Eq.~\eqref{eq:errorprop}, we may write 
${\rm Var}(\hat{\Omega})=\hat{\Omega}^2[{\rm Var}(\hat{h})/\hat{h}^2+{\rm Var}(\hat{\lambda})/\hat{\lambda}^2 -2 {\rm Cov}(\hat{h},\hat{\lambda})/\hat{h}\hat{\lambda} ]$. Taking the putative values of ${\rm Var}(\hat{h})$, ${\rm Var}(\hat{\lambda})$ and ${\rm Cov}(\hat{h},\hat{\lambda})$ to be the elements of the QFIM pseudoinverse $(\mathcal{M I}_{\lambda,h})^+$, we can therefore obtain an estimate for ${\rm Var}(\hat{\Omega})$. This provides a direct means of comparing the true precision bound from the effective QFI, with that obtained by the pseudoinverse method. The pseudoinverse of the QFIM Eq.~\eqref{eq:xy_lam_h_qfim} yields,
\begin{equation}
    \label{eq:pse}
    (\hat{\mathcal{I}}_{\lambda,h}^{\rm ring})^+ = \left(\sum_{k}\frac{16\sin^2(k)}{(\epsilon_k)^4}\right)^{-1}\times \frac{1}{\gamma^2(h^2+\lambda^2)^2}
    \begin{pmatrix}
   \; h^2 \;\; & -h \lambda \;\\
   \; -h \lambda \;\; & \lambda^2 \; 
\end{pmatrix}\;.
\end{equation}
Using this with the error propagation formula, we find by direct calculation that ${\rm Var}(\hat{\Omega})$ obtained using the pseudoinverse is precisely equal to the exact result obtained from the CRB via $ \mathcal{I}_{\Omega}$. Of course, one must already know the functional form of the effective parameter $\Omega=f(\theta_1,\theta_2)$ in order to apply the error propagation formula to obtain this result.

The fact that the QFIM pseudoinverse can reproduce the CRB for the effective parameter is a general result. Starting from the singular QFIM in the original coordinates $\hat{\mathcal{I}}(\vec{\theta})$, one can re-parametrize via Eq.~\eqref{eq:Ireparam} to obtain $\hat{\mathcal{I}}(\vec{\chi})=(\hat{M}^T)^{-1}\hat{\mathcal{I}}(\vec{\theta})\hat{M}^{-1}$, where $\vec{\chi}=(\Omega,\Lambda)$. The transformation $\hat{M}$ is chosen such that the only non-zero element is $[\hat{\mathcal{I}}(\vec{\chi})]_{\Omega,\Omega}\equiv \mathcal{I}_{\Omega}$. Now, one cannot invert the full matrix $\hat{\mathcal{I}}(\vec{\chi})$ but we do not need to: the CRB applies to the non-zero element ${\rm Var}(\hat{\Omega})=(\mathcal{M I}_{\Omega})^{-1}$. Alternatively, one can take the pseudoinverse $[\hat{\mathcal{I}}(\vec{\chi})]^+$, which is well-defined and whose only non-zero element is $1/\mathcal{I}_{\Omega}$. Taking ${\rm Var}(\hat{\Omega})=[\mathcal{M}\hat{\mathcal{I}}(\vec{\chi})]^+_{\Omega,\Omega}$ we get the correct result. 

This also works in terms of the original coordinates, since $[\mathcal{M}\hat{\mathcal{I}}(\vec{\chi})]^+= [(\hat{M}^T)^{-1}\hat{\mathcal{I}}(\vec{\theta})\hat{M}^{-1}]^{+}=\hat{M}[\hat{\mathcal{I}}(\vec{\theta})]^+ \hat{M}^T$. Taking\\ ${\rm Cov}(\vec{\theta})\ge [\hat{\mathcal{I}}(\vec{\theta})]^+$ yields $\hat{M} {\rm Cov}(\vec{\theta})\hat{M}^T \ge [\mathcal{M}\hat{\mathcal{I}}(\vec{\chi})]^+$ which from the error propagation formula Eq.~\eqref{eq:errorprop} gives precisely ${\rm Cov}(\vec{\chi})$ on the left hand side as desired.

This shows that the pseudoinverse correctly contains information about the effective CRB, once transformed to the correct basis of effective parameters.


\section{Lifting QFIM singularities through thermal fluctuations}\label{sec:lift_sing} 

As discussed in the previous sections,  QFIM singularities are a direct consequence of over-parametrization, which in turn leads to the emergence of metrological symmetries. To lift the QFIM singularity, one has to break the metrological symmetries. One route is to do this by  adding external (known) controls on the system. As an example, in Ref.~\cite{mihailescu2024critical} the singularity is lifted by applying an external local field. When singularities arise due to partial accessibility of a probe, another approach to lift singularities is to simply expand the support of the reduced state used for estimation. This was illustrated in Fig.~\ref{fig:heisenberg_post} for the Heisenberg chain model.

\begin{figure}[t]
    \centering
    \includegraphics[width=0.8\linewidth]{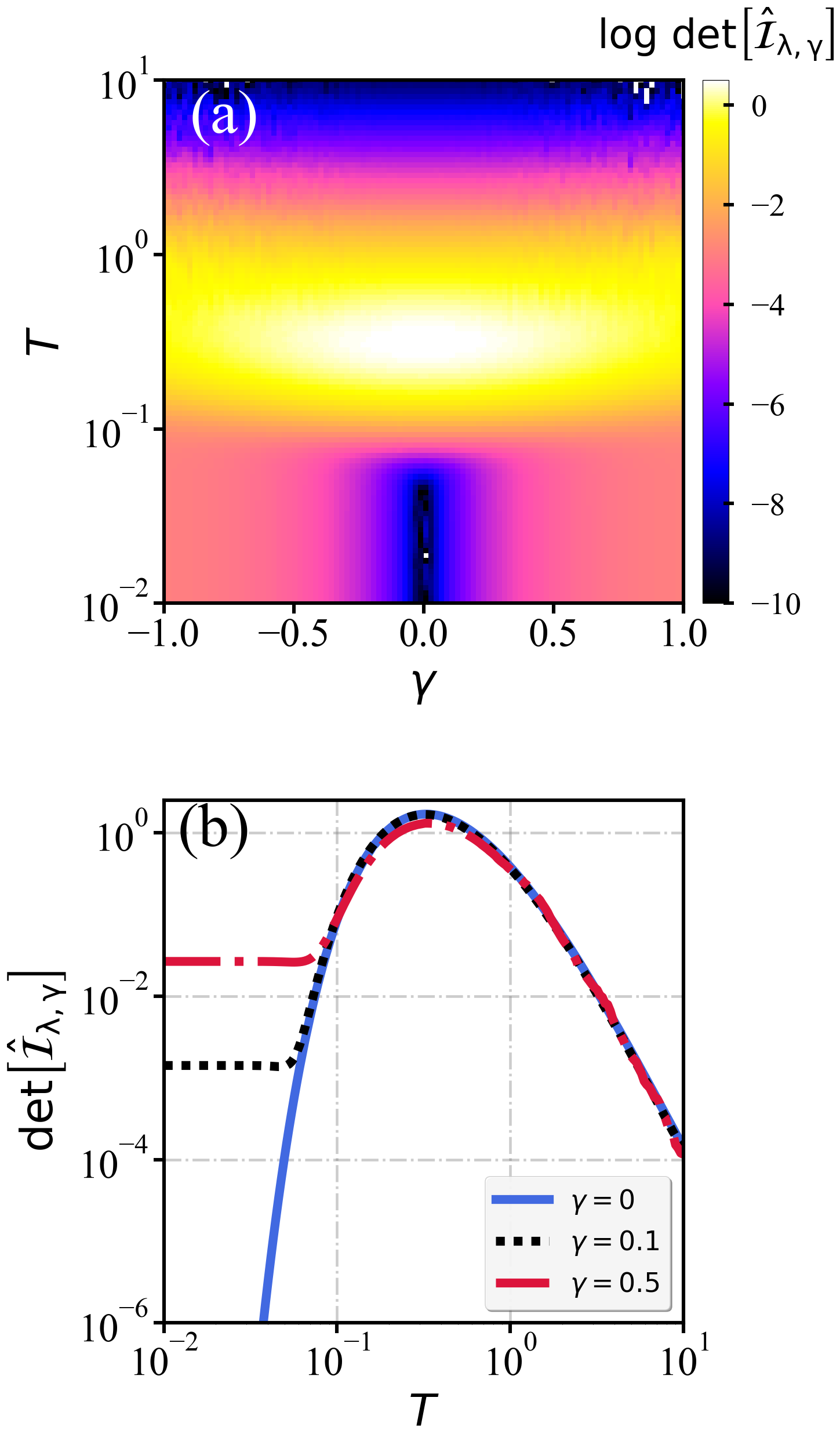}
\caption{\textbf{Multi-parameter estimation at finite temperatures.}  For the $XY$ model on the ring geometry, we compute the determinant of the QFIM for joint estimation of $\lambda$ and $\gamma$ at finite $T$. (a) Metrological phase diagram, showing $\text{det}[ \hat{I}_{\lambda,\gamma}]$ as a function of $\gamma$ and $T$ on a log colourscale. (b) $\text{det}[ \hat{I}_{\lambda,\gamma}]$ vs $T$ for $\gamma=0$, $0.1$ and $0.5$, showing low-temperature suppression near the singular point at $\gamma=T=0$. Throughout we use $N=4$ and set $h=1$ and $\lambda=0.6$.}
    \label{fig:xy_temperature_effect}
\end{figure}

In this section, we present an alternative. We show below that \textit{thermal fluctuations}, which are typically considered as having a destructive effect on quantum resources, can often be leveraged to lift QFIM singularities. Therefore, perhaps counterintuitively, metrological utility is actually \textit{enhanced} at elevated temperatures in such systems. It was recently demonstrated that, under certain circumstances, small temperature fluctuations can enhance the QFI for single-parameter estimation~\cite{abiuso2025fundamental,PhysRevA.109.L050601}, and our findings extend this to the multi-parameter context. In fact, another way of viewing our result is that the existence of QFIM singularities causes a \textit{suppression} of the \textit{low-temperature} metrological precision. Therefore, even for non-singular systems, the existence of a singular point nearby in the phase diagram can cause partial suppression of sensitivity, and these are precisely the systems where thermal fluctuations can boost quantum sensing.

We illustrate these concepts by considering the simultaneous estimation of the anisotropy $\gamma$ and coupling $\lambda$ in the $XY$ model for $N=4$ spins on the ring geometry, see Eq.~\eqref{eq:xy_general_ham}. As before, we assume the field is known and set $h = 1$. 
When any pure state of the system is used for sensing, the QFIM is singular when $\gamma=0$ (see Appendix~\ref{app:xydet} and Eq.~\eqref{eq:xy_ham_gam_det_n4}). However, the QFIM of the thermal state at finite $T$ is \textit{non-singular}.

On the other hand, at sufficiently low temperatures $T\ll \Delta$ the system is described by its ground state, where $\Delta$ is the energy gap between the ground and first-excited state. The thermal population of excited states is exponentially suppressed in this limit, and so one expects the QFIM to be determined by Eq.~\eqref{eq:Pure_State_QFI} for the ground state. This is demonstrated explicitly in Appendix \ref{app:temp}, where the QFIM for a general thermal state is shown to be continuously connected to the ground state QFIM by taking the limit $T\to 0$. 
This implies that the determinant of the QFIM (and hence the overall precision of multi-parameter estimation which is proportional to the determinant) continuously vanishes as $T\to 0$ for singular systems. 
In fact, the existence of a singularity in the ground state QFIM can be shown to produce an exponentially-strong suppression at small finite temperatures, $\text{det}[ \hat{\mathcal{I}}] \sim e^{-\Delta/T}$. 

This is demonstrated in Fig.~\ref{fig:xy_temperature_effect} where we compute the  QFIM determinant $\text{det} [ \hat{\mathcal{I}}_{\lambda,\gamma}^{\text{ring}} ]$
for joint estimation of $\lambda$ and $\gamma$ in the $XY$ model on the ring geometry. In panel (a) we show the metrological phase diagram as a function of $\gamma$ and $T$ for representative $\lambda=0.6$ and $h=1$. The QFIM singularity is seen as a low-temperature suppression of the determinant near $\gamma=0$ (note the log colourscale). This is because the pure state QFIM for the ground state at $\gamma=0$ is singular and so $\text{det} [ \hat{\mathcal{I}}_{\lambda,\gamma}^{\text{ring}} ]=0$ at $\gamma=T=0$. However, the broader influence of this singular point in the phase diagram is clearly evident. By continuity, the low-temperature properties of systems with small finite $\gamma$ also show a low-temperature suppression of the QFIM determinant, although not all the way to zero. This is shown more clearly by the line plots in Fig.~\ref{fig:xy_temperature_effect}(b).

In the opposite limit, $T \!\to\! \infty$, the thermal state becomes maximally mixed and no information about the parameters is encoded. Therefore the QFIM determinant vanishes (for all values of $\gamma$ or $\lambda$). As a consequence, the QFIM determinant must \textit{peak} at finite $T$, indicating that optimal sensitivity is indeed found at intermediate temperatures.

We conclude that, in the $XY$ model, at all finite temperatures the QFIM singularity is lifted. This means that the QFIM is invertible, the CRB holds, and the BVM theorem implies that Bayesian estimation will converge to a unique point in parameter space, corresponding to the true values of the parameters to be estimated. However, we note that close to a singularity, when the QFIM determinant is finite but small, the rate of convergence will be extremely slow, and long-lived signatures of the singular lines in the Bayesian posterior distributions are expected (albeit that these must eventually evaporate). 

We therefore emphasize that the fundamental categorization of the ground state QFIM as singular or non-singular is highly important.  Indeed, as shown in Fig.~\ref{fig:xy_temperature_effect}, it is also useful to know when a system is close to, but not at, a point in the phase diagram where the ground state QFIM is singular. This points to metrological maps of the ground state QFIM determinant as being a particularly useful diagnostic tool. 

Finally, we note that some systems may be singular also at finite temperatures -- see for example the singular reduced thermal state of the Heisenberg chain in Sec.~\ref{sec:heisenberg}. This underlines the importance of determining the physical origin of the QFIM singularity, and whether metrological symmetries exist only at the level of pure states, or survive for mixed states. We have provided examples of both in this paper.


\section{Conclusions}

Quantum sensing lies at the foundation of all emerging quantum technologies, with applications spanning from device calibration and readout optimization to the estimation of fundamental constants of the universe, as well as the precise measurement of external electromagnetic or gravitational fields. The Cram\'{e}r-Rao formalism provides theoretical bounds on achievable precisions for a given quantum probe.  For single-parameter quantum sensing, in which all properties of a system are assumed to be known except for the one parameter being estimated, this formalism provides a full characterization. It gives a tight bound for the ultimate obtainable precision and a systematic strategy to achieve it. In the case of multi-parameter quantum sensing however, the situation becomes far more complex because the Cram\'{e}r-Rao approach involves a \textit{matrix} inequality, which features the inverse of the CFIM for a given measurement, or the inverse of the QFIM for an optimized measurement. Two major issues therefore arise when applying the Cram\'{e}r-Rao formalism to multi-parameter problems. First, the bound given by the QFIM is in general no longer tight due to the well-studied measurement incompatibility issue. Second, the Fisher information matrix might be singular and thus its inverse does not exist. In many cases, the CFIM for a particular measurement might be singular, even though the QFIM for the optimal measurement is invertible. In such a scenario, the CFIM singularity is related to incomplete measurement outcomes and can be overcome by adopting a better measurement strategy. The QFIM singularity, however, is more consequential as it directly implies the singularity of the CFIM for all possible measurements. 

In this paper, we develop a framework to understand the different types of QFIM singularities and their physical origin. Furthermore, we show how Bayesian estimation  can provide detailed and highly non-trivial information about singular multi-parameter problems, including on the effective parameter encoding and the existence of metrological symmetries. Utilizing this information, which requires no \textit{a priori} knowledge of the system, we demonstrate that a well-defined and non-singular estimation problem can be obtained by suitable re-parametrization (which is in general a \textit{non-linear} coordinate transformation). The effective parameters can then be estimated with a precision that follows from an effective CRB.  

We show that the key to understanding QFIM singularities is an intrinsic over-parametrization on the metrological level. This is related to the emergence of metrological symmetries: systems with different parameters cannot be distinguished by measurements on the system and asymptotically yield the same statistical model from which inferences about the parameters are made.  
Such an over-parametrization can arise when the quantum state or the optimal measurement operators depend on the $d$ bare parameters to be estimated, $\vec{\theta}$, only through a \textit{reduced set} of $k<d$ effective parameters, $\vec{\Omega}$. However, singularities can also result from ``hidden'' effective parameters that are not apparent on the level of the state or measurement operators. In all cases, the metrological over-parametrization and hence the QFIM singularity is signalled by one or more zero eigenvalues of the QFIM. The number of non-zero eigenvalues is the number of effective parameters which can be estimated. 

Although analysis of the QFIM itself does not typically yield information about the functional dependence of the effective parameters or the nature of the metrological symmetries in singular systems, this information is readily accessible from the Bayesian posterior distribution. Indeed, we note that this information can be extracted by post-processing the data: the standard Bayesian strategy in terms of the original parameters can still be used without modification for the purposes of data collection and the experimental protocol.

We also point out that metrological `phase diagrams' of the QFIM determinant are particularly useful diagnostic tools. The topology of these phase diagrams is controlled by the singular systems with zero determinant, and understanding when the ground state of a system is singular can be rather informative. In particular, we find that the measurement precision for non-singular systems is strongly affected by singular points nearby in the phase diagram.

We believe that our results offer an insight into the fundamental origin of metrological singularities, and our approach for identifying effective parameters and metrological symmetries through Bayesian analysis provides a systematic route to estimation even when the standard frequentist approach fails.

\begin{acknowledgments}
We acknowledge insightful discussions with Karol Gietka. GM acknowledges support from Equal1 Laboratories Ireland Limited. AKM acknowledges support from Science Foundation Ireland through Grant 21/RP-2TF/10019. SC acknowledges support from the John Templeton Foundation Grant ID 62422 and the Alexander von Humboldt Foundation. SS acknowledges support from National Natural Science Foundation of China (Grant No.~W2433012). AB acknowledges support from National Natural Science Foundation of China (Grants Nos.~12050410253, 92065115, and 12274059) and the Ministry of Science and Technology of China (Grant No.~QNJ2021167001L).
\end{acknowledgments}

\appendix

\section{Illustrative example -- Heisenberg trimer}\label{app:triangle}
Here we provide a simple but nontrivial example of a QFIM singularity, results for which we presented for illustrative purposes in Fig.~\ref{fig:illustration}. 
We consider a triangular arrangement of three exchange-coupled spins-$\tfrac{1}{2}$, which admits an exact solution for the QFIM. Our model reads,
\begin{eqnarray}\label{eq:Htriangle}
    \hat{H}&=&K\vec{\boldsymbol{S}}_1\cdot\vec{\boldsymbol{S}}_2 + J(\vec{\boldsymbol{S}}_1\cdot\vec{\boldsymbol{S}}_3+\vec{\boldsymbol{S}}_2\cdot\vec{\boldsymbol{S}}_3) \nonumber \\&\equiv& E_0+\tfrac{1}{2}J\vec{\boldsymbol{S}}^{\:2}_{\rm tot} + \tfrac{1}{2}(K-J)\vec{\boldsymbol{S}}_{12}^{\:2}\;, 
\end{eqnarray}
where on the second line we used $\vec{\boldsymbol{S}}_{\rm tot}=\vec{\boldsymbol{S}}_1+\vec{\boldsymbol{S}}_2+\vec{\boldsymbol{S}}_3$ and $\vec{\boldsymbol{S}}_{12}=\vec{\boldsymbol{S}}_1+\vec{\boldsymbol{S}}_2$. Here $E_0=-\tfrac{3}{8}J-\tfrac{3}{4}K$ is an irrelevant constant. Since $[\hat{H},\vec{\boldsymbol{S}}_{\rm tot}^{\:2}]=[\hat{H},\vec{\boldsymbol{S}}_{12}^{\:2}]=[\hat{H},\hat{S}^z_{\rm tot}]=[\vec{\boldsymbol{S}}_{\rm tot}^{\:2},\vec{\boldsymbol{S}}_{12}^{\:2}]=[\vec{\boldsymbol{S}}_{\rm tot}^{\:2},\hat{S}^z_{\rm tot}]=0$, all eigenstates $|S_{\rm tot},S_{12},S_{\rm tot}^z\rangle$ of $\hat{H}$ are uniquely identified by the quantum numbers $S_{\rm tot}$, $S_{12}$ and $S_{\rm tot}^z$. The energies of these states are determined just by $S_{\rm tot}$ and $S_{12}$ according to Eq.~\eqref{eq:Htriangle}, $E_{S_{\rm tot},S_{12}}=E_0+\tfrac{1}{2}JS_{\rm tot}(S_{\rm tot}+1)+\tfrac{1}{2}(K-J)S_{12}(S_{12}+1)$.
We therefore have a 2-fold degenerate doublet state with $S_{12}=0$ and energy $E_{1/2,0}=-\tfrac{3}{4}K$, a 2-fold degenerate doublet state with $S_{12}=1$ and energy $E_{1/2,1}=\tfrac{1}{4}K-J$, and a 4-fold degenerate quartet state with $S_{12}=1$ and energy $E_{3/2,1}=\tfrac{1}{4}K+\tfrac{1}{2}J$. The thermal weights (density matrix probabilities) are given by $p_{S_{\rm tot},S_{12}}=e^{-E_{S_{\rm tot},S_{12}}/T}/Z$ with partition function $Z=2e^{-E_{1/2,0}/T}+2e^{-E_{1/2,1}/T}+4e^{-E_{3/2,1}/T}$. Since the symmetry quantum numbers fully determine all states, the basis is fixed. The (thermal state) QFIM then follows as~\cite{liu2019quantum}, $\hat{\mathcal{I}}_{i,j}=2(\partial_i p_{1/2,0}\:\partial_j p_{1/2,0})/p_{1/2,0}+2(\partial_i p_{1/2,1}\:\partial_j p_{1/2,1})/p_{1/2,1}+4(\partial_i p_{3/2,1}\:\partial_j p_{3/2,1})/p_{3/2,1}$. 
The QFIM for two-parameter estimation of any pair $(K,T)$ or $(K,J)$ or $(J,T)$ is nonsingular.

We now consider the reduced state of spins 1 and 2 after tracing out spin 3. This is trivially computed because by SU(2) spin symmetry, the reduced states can be labelled by their spin $S_{12}$ but the original full states are also labelled by $S_{12}$. It is straightforwardly shown that for the reduced states we have a 3-fold degenerate triplet with $S_{12}=1$ and probability $p^{\rm red}_{1}=\tfrac{2}{3}p_{1/2,1}+\tfrac{4}{3}p_{3/2,1}$ as well as a unique singlet $S_{12}=0$ with probability $p^{\rm red}_{0}=2p_{1/2,0}$. The QFIM for the reduced (thermal) state is then $\hat{\mathcal{I}}^{\rm red}_{i,j}=3(\partial_i p^{\rm red}_{1}\:\partial_j p^{\rm red}_{1})/p^{\rm red}_{1}+(\partial_i p^{\rm red}_{0}\:\partial_j p^{\rm red}_{0})/p^{\rm red}_{0}$. With this, it follows that the reduced state QFIM is singular for any $T$, with $\rm{det}[\hat{\mathcal{I}}^{\rm red}]=0$. 

This can be understood as an example of singularity due to state dependence on an effective implicit parameter. The reduced density matrix $\hat{\varrho}_{\rm red}(\Omega)$ is controlled entirely by the thermal expectation value $\Omega\equiv \langle \vec{\boldsymbol{S}}_{12}^{\:2} \rangle = 6p^{\rm red}_{1}$ since $3p^{\rm red}_{1}+p^{\rm red}_{0}=1$. Thus from Eq.~\eqref{eq:effective_param_form} we immediately conclude that the QFIM is singular. In this case, the result follows because the reduced state only depends on $J$, $K$ and $T$ through the effective parameter $\Omega\equiv \Omega(J,K,T)$. Measurements on the system cannot distinguish between different parameter combinations of $J$, $K$ and $T$ if they give the same value of $\Omega$. 

\begin{figure}[t]
\centering
    \includegraphics[width=0.36\textwidth]{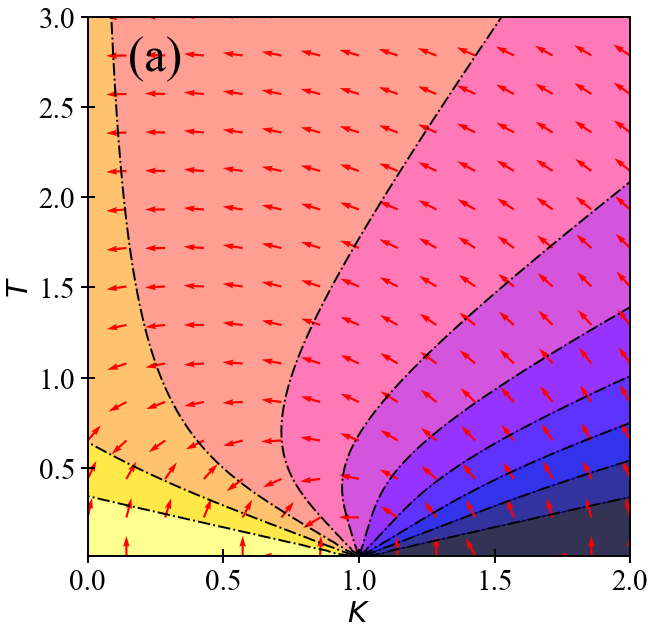}
    \hfill
    \includegraphics[width=0.36\textwidth]{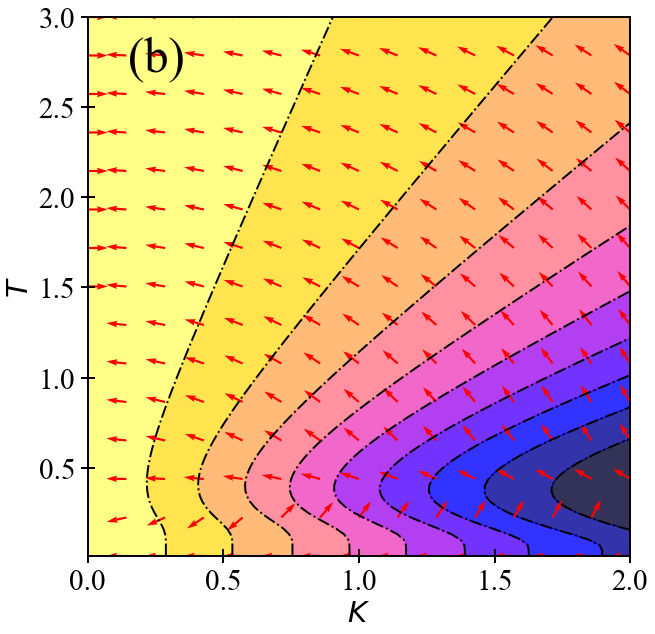}
	\caption{(a) Contour lines of constant $\Omega= \langle \vec{\boldsymbol{S}}_{12}^{\:2} \rangle$ for the Heisenberg trimer model considered in Fig.~\ref{fig:illustration}. (b) Contour lines of constant $\Omega=\langle \hat{\vec{\bm{\sigma}}}_{\text{2}} \cdot \hat{\vec{\bm{\sigma}}}_{\text{3}} \rangle$ for the Heisenberg chain model with $N=4$ sites considered in Sec.~\ref{sec:heisenberg}. We set $J = 1$ in both cases. All Bayesian posterior lines in the singular scenario must converge to one of these contour lines. Overlaid as the red arrows in both cases are the eigenvector fields corresponding to the single non-zero QFIM eigenvalue.}
	\label{fig:Contour_Corr}
\end{figure}

In the context of Bayesian estimation, the posterior will therefore converge to a contour line of constant $\Omega$, passing through the true parameter values. In the Heisenberg trimer example, we can find an expression for these contour lines analytically, $K=T \ln[(2-\Omega) e^{-J/2 T} (2 + e^{3 J/2 T})/\Omega]$. These contour lines are plotted in Fig.~\ref{fig:Contour_Corr}(a). Overlaid on the figure as the red arrows is the QFIM eigenvector field corresponding to the single non-zero QFIM eigenvalue, confirming Eq.~\eqref{eq:evect_field}. We emphasize that it is not possible to analytically extract the functional dependence of $\Omega$ on the bare parameters $K$, $J$ and $T$ from the QFIM eigenvectors by inverting Eq.~\eqref{eq:evect_field} even in this simple case. The numerical solution is also very challenging, being plagued by divergences and branch cuts. On the other hand, Bayesian estimation strategies are straightforward.

In the lower panels of Fig.~\ref{fig:illustration} we consider multiparameter Bayesian estimation using the (singular) reduced state of the Heisenberg trimer model, for the unknown parameters $\theta_1=K$ and $\theta_2=T$, showing results for the Bayesian posterior of the temperature estimator after $\mathcal{M}=100$, $300$ and $1000$ measurements with representative $J=1$. The true value of the parameters is $T^{\rm tr}=1$ and $K^{\rm tr}=0.5$ which yields $\Omega^{\rm tr}\simeq 1.4$. The Bayesian posterior is indeed found to converge precisely to our exact expression for the contour line corresponding to $\Omega\simeq 1.4$. This shows vividly that the effective metrological symmetries revealed by the Bayesian estimation strategy can be highly nontrivial and can have complicated functional dependences on the underlying system parameters.

In the upper panels of Fig.~\ref{fig:illustration}, we show a non-singular example for comparison. Here the singularity of the Heisenberg trimer reduced state is lifted by application of a Zeeman field $B\hat{S}_{\rm tot}^z$, which breaks the $S_{12}=1$ multiplet degeneracy of the reduced state. In this case the BVM theorem holds and we expect to see convergence of the Bayesian posterior to a unique point. This is confirmed by our results for $B=1$.


\section{Solution of the XY model on the ring geometry}\label{app:xydet}
We consider here the XY model Eq.~\ref{eq:xy_general_ham} on the ring geometry. In this case the system can be mapped to a non-interacting fermionic model via the Jordan-Wigner transformation. The model is then Fourier transformed to momentum space and solved exactly \cite{pfeuty1970the, barouch1970statistical,10.21468/SciPostPhysLectNotes.82} using fermionic two-component spinors $\hat{\Psi}^{\dagger}_{k} = \left( \hat{c}_k^{\phantom{\dagger}}, \hat{c}_{-k}^{\dagger} \right)$, 
\begin{equation}
    \label{eq:xy_ham_pseudo_basis}
    \hat{H}_{k} = 2 \Psi^{\dagger}_k \left[ \left(h + \lambda \cos{k}\right)\hat{\sigma}_z + \left( \gamma \lambda \sin{k} \right)  \hat{\sigma}_x \right] \Psi^{\phantom{\dagger}}_k\;,
\end{equation}
which decouples the Hamiltonian as $\hat{H}_{XY}^{\rm ring} = \sum_{k>0} \hat{\Psi}^\dagger_k \hat{H}_{k} \hat{\Psi}^{\phantom{\dagger}}_k$. Solving the eigenvalue problem for Eq.~\eqref{eq:xy_ham_pseudo_basis} yields the dispersion for each band, $\epsilon_k^{\pm} = \pm 2 \sqrt{\left(h + \lambda \cos{k} \right)^2 + \lambda^2\gamma^2 \sin^2{k}}$. In each momentum sector $k$, the ground state energy is  $\epsilon_k^-$ and the corresponding ground state is $\ket{\psi_{GS}}_{k} = \cos{\frac{\theta_k}{2}} \ket{0}_k + \sin{\frac{\theta_k}{2}} \ket{1}_k$ with $\tan{\theta_k} = -\frac{\lambda \gamma \sin{k}}{h + \lambda \cos{k}}$. The full ground state for an $N$-site system is therefore given as the tensor product of those in each momentum-space block, $\ket{\psi_{GS}}_N = \bigotimes_{k>0}\ket{\psi_{GS}}_k$. 

For joint estimation of $\lambda$ and $\gamma$ the QFIM for arbitrary $N$ can be found analytically,
\begin{equation}
    \label{eq:xy_gam_lam_qfim_app}
    \hat{\mathcal{I}}_{\lambda, \gamma}^{\rm ring} = \sum_{k}\frac{16\sin^2(k)}{(\epsilon_k^-)^4}
    \begin{pmatrix}
    h^2 \gamma^2 \;\; & h \gamma \lambda \left( h + \lambda \cos{k} \right)  \\ h \gamma \lambda \left( h + \lambda \cos{k} \right)\;\; &  \lambda^2\left(h + \lambda \cos{k}\right)^2
\end{pmatrix}\;,
\end{equation}
with $k =  \pi(2n+1)/N$ and $n = 0,1,2,\ldots,\left\lfloor\frac{N}{2}\right\rfloor - 1$ labelling the momentum space blocks. 
From this, we can construct the determinant of the QFIM, which takes a simple analytic form for $N=4$ spins, 
\begin{equation}
    \label{eq:xy_ham_gam_det_n4}
    \text{det}\left[\hat{\mathcal{I}}^{\rm ring}_{\lambda,\gamma} \right] = \frac{8 h^2 \gamma^2 \lambda^4}{\left(4 h^2 + 4 h^2 \left[\gamma^2 - 1\right] \lambda^2 + \left[\gamma^2 + 1 \right]^2 \lambda^4\right)^2}\;.
\end{equation}
We deduce that the QFIM for the XY model on the ring geometry for joint estimation of $\lambda$ and $\gamma$ is singular only when $h$, $\gamma$ and/or $\lambda$ are zero.


\section{When is the pure-state QFIM of the\\two-level system singular?}\label{app:2ls}
\noindent Consider the most general pure state for a two-level system,
\begin{equation}\label{eq:2ls}
    |\psi\rangle = \cos[p(\vec{\theta})]|0\rangle + \sin[p(\vec{\theta})]\:{\rm exp}[i q(\vec{\theta})]|1\rangle \;,
\end{equation}
where $p(\vec{\theta})$ and $q(\vec{\theta})$ are arbitrary real functions of the parameters $\vec{\theta}=(\theta_1,\theta_2,\theta_3,...)$ to be estimated.  Elements of the QFIM from Eq.~\eqref{eq:Pure_State_QFI} in this case are simply given by,
\begin{equation}
\hat{\mathcal{I}}_{i,j}=4\partial_{\theta_i}p(\vec{\theta})\partial_{\theta_j}p(\vec{\theta})+\partial_{\theta_i}q(\vec{\theta})\partial_{\theta_j}q(\vec{\theta})\sin^2[2p(\vec{\theta})] \;.
\end{equation}

From this we first note that estimation of $d=3$ or more parameters with a two-level system is impossible. This is a special case of the result in \cite{Candeloro_2024}, and here one can verify that ${\rm det}[\hat{\mathcal{I}}]=0$ irrespective of the functions $p(\vec{\theta})$ and $q(\vec{\theta})$ for $d\ge 3$. 
For the two-parameter estimation problem where $\vec{\theta}=(\theta_1,\theta_2)$ we can ask: when is the QFIM of this pure state singular? 
Assuming finite real functions $p$ and $q$, it is straightforward to show that the determinant of the QFIM for the state Eq.~\eqref{eq:2ls} is exactly zero if and only if, 
\begin{equation}\label{eq:2ls_singular}
    \{p,q\}_{\vec{\theta}}\equiv \partial_{\theta_1}p(\vec{\theta})\partial_{\theta_2}q(\vec{\theta})-\partial_{\theta_2}p(\vec{\theta})\partial_{\theta_1}q(\vec{\theta})=0 \;.
\end{equation}

If the state depends only on a single function of the bare parameters  $\Omega(\theta_1,\theta_2)$ then this implies from Eq.~\eqref{eq:2ls} that $p(\Omega)$ and $q(\Omega)$ are both functions of $\Omega$. In this case the QFIM is clearly always singular from the condition Eq.~\eqref{eq:2ls_singular} (and in agreement with Eq.~\eqref{eq:effective_param_form}).

Note however that the QFIM can also be singular even with different functions $p(\vec{\theta})$ and $q(\vec{\theta})$, such that the state depends separately on both parameters $\theta_1$ and $\theta_2$ and not just on a single function $\Omega(\theta_1,\theta_2)$. In such a case, the Bayesian posterior distribution can still reveal the hidden metrological symmetry encoded in an effective parameter $\Omega(\theta_1,\theta_2)$.

Finally, we comment that for mixed states of the two-level system, the conditions for QFIM singularity are more inscrutable. However, the Bayesian approach can still yield useful information for singular mixed states (except for maximally-mixed states which contain no information about the underlying parameters).


\section{Contour plots of constant $\Omega$\\in the Heisenberg chain model}\label{app:heis_chain}
In Sec.~\ref{sec:heisenberg} we consider the Heisenberg spin chain model with $N=4$ sites. The reduced state of the central two spins has a singular QFIM. However, unlike the Heisenberg trimer example considered above, in this case there is no analytic expression for the state, QFIM, or contour lines of the correlator $\Omega=\langle \hat{\vec{\bm{\sigma}}}_{\text{2}} \cdot \hat{\vec{\bm{\sigma}}}_{\text{3}} \rangle$ which controls the reduced density matrix. The lines of likelihood in the Bayesian posterior must still converge to a contour line of constant $\Omega$. These are numerically computed for the Heisenberg chain model and shown in Fig.~\ref{fig:Contour_Corr}(b). Bayesian estimation can therefore still be used to uncover the state dependence on $\Omega$ and the underlying dependence of $\Omega$ on the system parameters, as shown in Fig.~\ref{fig:heisenberg_post}. 

Overlaid on Fig.~\ref{fig:Contour_Corr}(b) as the red arrows is the QFIM eigenvector field corresponding to the single non-zero QFIM eigenvalue, confirming Eq.~\eqref{eq:evect_field}.


\section{XY model QFIM for $N = 3$}\label{app:xyn3}
For $N=3$ the ring and all-to-all geometries of the XY model are of course equivalent. The corresponding QFIM for estimation of the parameters $\lambda$ and $\gamma$ can be computed exactly, and is given by,
\begin{equation}
    \label{eq:xy_n3}
    \hat{\mathcal{I}}_{\lambda,\gamma} = \frac{1}{\chi} \begin{pmatrix}
         12 \gamma ^2 h^2 & 6 \gamma  h \lambda  (2 h+\lambda ) \\ 6 \gamma  h \lambda  (2 h+\lambda ) & 3 \lambda ^2 (\lambda +2 h)^2  
    \end{pmatrix}
\end{equation}
where $ \chi = [\left(3 \gamma ^2+1\right) \lambda ^2+4 h^2+4 h \lambda]^2$. Clearly the QFIM is singular since its determinant exactly vanishes for all values of $\lambda$, $\gamma$ and $h$. 
For $N=3$ we have a single momentum space block in Eq.~\eqref{eq:xy_gam_lam_qfim_app} for $n=0$ corresponding to $k=\pi/3$. Thus the ground state $\ket{\psi_{GS}}_3$ used for quantum sensing is parametrized by a single angle $\theta_{\pi/3}$ defined by $\tan \theta_{\pi/3}=-\sqrt{3}\lambda\gamma/(2h+\lambda)$. We see immediately that for $N=3$ the state is controlled by a single effective parameter $\Omega=\lambda\gamma/(2h+\lambda)$ which is at heart responsible for rendering the multi-parameter estimation problem singular.

The QFIM has only one non-zero eigenvalue,
\begin{equation}
    G_1 = \frac{3 \left(4 h^2 \left(\gamma ^2+\lambda ^2\right)+4 h \lambda ^3+\lambda
   ^4\right)}{\left(\left(3 \gamma ^2+1\right) \lambda ^2+4 h^2+4 h \lambda
   \right)^2}
\end{equation}
with corresponding eigenvector given by,
\begin{equation}
    \vec{u}_1 = \begin{pmatrix}
        \frac{2 \gamma  h}{\sqrt{4h\lambda^3+\lambda^4+4h^2(\gamma^2+\lambda^2)}} \\ \frac{1}{\sqrt{\frac{4 \gamma ^2
   h^2}{\lambda ^2 (\lambda +2 h)^2}+1}}
    \end{pmatrix}
\end{equation}
The eigenvector corresponding to the zero eigenvalue $G_2 = 0$ is given by,
\begin{equation}
    \vec{u}_2 = \begin{pmatrix}
        \frac{-\lambda  (2 h+\lambda )}{\sqrt{4h\lambda^3+\lambda^4+4h^2(\gamma^2+\lambda^2)}} \\ \frac{1}{\sqrt{\frac{\lambda ^2 (2
   h+\lambda )^2}{4 \gamma ^2 h^2}+1}}
    \end{pmatrix}
    \end{equation}

In this case, one can directly verify Eqs.~\eqref{eq:qfim_eval} and \eqref{eq:evect_field} for the effective parameter $\Omega=\lambda\gamma/(2h+\lambda)$. Using Eq.~\eqref{eq:Ireparam} we readily obtain,
\begin{align}
    \mathcal{I}(\Omega) = \frac{3 (2h+\lambda)^4}{(4h^2 + 4h\lambda + [1+3\gamma^2]\lambda^2)^2} \;.
\end{align}

Note that for $N>3$ we have involvement of multiple momentum-space blocks and therefore the ground state is certainly \textit{not} controlled by just a single effective parameter.


\section{Recovery of pure-state QFIM in the limit $T\to 0$}\label{app:temp}
Elements of the QFIM for general mixed states are given by Eq.~\eqref{eq:QFIM_Entry}, whereas for pure states the QFIM elements are given by Eq.~\eqref{eq:Pure_State_QFI}. When $\hat{\varrho}\equiv \hat{\varrho}(T)$ describes a \textit{thermal} state, the well-controlled $T\to 0$ (zero temperature) limit of Eq.~\eqref{eq:QFIM_Entry} should yield Eq.~\eqref{eq:Pure_State_QFI} for the \textit{ground state}, with $\hat{\varrho}= | \psi_{\text{GS}} \rangle \langle \psi_{\text{GS}}|$. This is important to establish continuity between idealized results for pure states and those for general thermal states. We show that explicitly here. 

For a thermal state $\hat{\varrho}$ we use the spectral decomposition $\hat{\varrho}=\sum_j e_j |e_j\rangle\langle e_j|$, where the probabilities $e_j$ are the normalized Boltzmann weights $e_j=e^{-E_j/T}/Z$, with $Z=\sum_j e^{-E_j/T}$ the partition function and $E_j$ the energy of state $|e_j\rangle$ satisfying the Schr\"odinger equation $\hat{H}|e_j\rangle = E_j |e_j\rangle$. 

When the spectrum of the density matrix is strictly non-degenerate, we may employ the standard results for the derivatives of an eigensystem, $\partial_\mu e_j = \langle e_j | \partial_\mu \hat{\varrho} | e_j\rangle$ and $|\partial_\mu e_j\rangle = \sum_{i\ne j} \frac{\langle e_i | \partial_\mu \hat{\varrho} | e_j\rangle}{e_j-e_i} \times |e_i\rangle$ to write Eq.~\eqref{eq:QFIM_Entry} in the alternative form~\cite{liu2019quantum},
\begin{align}\label{eq:QFIM_alt}
\hat{\mathcal{I}}_{i,j} = \sum_n \frac{(\partial_i e_n)(\partial_j e_n)}{e_n} +2\sum_{n \ne m}\frac{(e_n-e_m)^2}{e_n+e_m}{\rm Re}~ \langle e_n| \partial_{i} e_m\rangle \langle \partial_j e_m|e_n\rangle  \;.
\end{align}
Defining now the ground state $|\psi_{\rm GS}\rangle$ as the state with lowest energy $E_{\rm GS}$, we specify all energies with respect to the ground state, $\Tilde{E}_j = E_j - E_{\text{GS}}$, and factorize the partition function as $Z=e^{-E_{\text{GS}} / T} \times \Tilde{Z}$ where $\Tilde{Z}=\sum_j e^{-\Tilde{E}_j / T}$. The thermal weights can then be expressed as $e_j=e^{-\Tilde{E}_j/T}/\Tilde{Z}$. This rewriting simplifies taking the $T\to 0$ limit since the factors $e^{-\Tilde{E}_j/T}$ tend to either $0$ or $1$ in a well-behaved fashion. Specifically, $e^{-\Tilde{E}_{\rm GS}/T}\to 1$ for the ground state with $\Tilde{E}_{\rm GS}=0$, and $e^{-\Tilde{E}_{\rm Ex}/T}\to 0$ for all excited states with $\Tilde{E}_{\rm Ex}>0$. Therefore $\Tilde{Z}\to 1$ as $T\to 0$ and so the thermal weights $e_{\rm GS}\to 1$ and $e_{\rm Ex}\to 0$.

The first term in Eq.~\eqref{eq:QFIM_alt} can be expanded as,
\begin{equation}\label{firstterm}
    \sum_n \frac{(\partial_i e_n)(\partial_j e_n)}{e_n}= \frac{\partial_{i} e_{\text{GS}} \partial_{j} e_{\text{GS}}}{e_{\text{GS}}} + \sum_{\text{Ex}}\frac{\partial_{i} e_{\text{Ex}} \partial_{j} e_{\text{Ex}}}{e_{\text{Ex}}}
\end{equation}
The first term for the ground state tends to $0$ as $T\to 0$ because the ground state thermal weights tend to $1$ independently of the parameters, and hence the derivatives on the numerator vanish. Although the excited state weights all tend to $0$ as $T\to 0$ and become parameter independent, the second term in Eq.~\eqref{firstterm} is more subtle due to the vanishing weight on the denominator. We therefore use L'Hopital's rule to write $\lim_{T\to 0}~(\partial_{i} e_{\text{Ex}} \partial_{j} e_{\text{Ex}})/e_{\text{Ex}} =\lim_{T\to 0} ~\partial_T(\partial_{i} e_{\text{Ex}} \partial_{j} e_{\text{Ex}})/\partial_T e_{\text{Ex}}$. Analysis of the latter shows that the sum over excited state contributions vanishes term by term as $T\to 0$.

This leaves only the second term in Eq.~\eqref{eq:QFIM_alt}. We distinguish two cases for the coefficient,
\begin{eqnarray}
    \sum_{n\ne m}\frac{(e_n-e_m)^2}{e_n+e_m} = \sum_{\substack{n\: \in \:{\rm Ex}\\ m\: \in\: \rm{Ex}}} \frac{(e_n-e_m)^2}{e_n+e_m} + 2\sum_{n \:\in\: {\rm Ex}} \frac{(e_{\rm GS}-e_n)^2}{e_{\rm GS}+e_n} \nonumber
\end{eqnarray}
noting that $n$ and $m$ cannot both be ground states because of the condition $n\ne m$ (and assuming a non-degenerate spectrum).
For the first term, we must use L'Hopital's rule again, $\lim_{T\to 0}~(e_n-e_m)^2/(e_n+e_m)=\lim_{T\to 0}~\partial_T(e_n-e_m)^2/\partial_T(e_n+e_m)=0$, so this contribution vanishes. For the second term $(e_{\rm GS}-e_{\rm Ex})^2/(e_{\rm GS}+e_{\rm Ex})\to 1$ as $T\to 0$.

Thus as $T\to 0$ we can write Eq.~\eqref{eq:QFIM_alt} as,
\begin{align}\label{eq:QFIM_alt2}
\begin{split}
\hat{\mathcal{I}}_{i,j} &=4{\rm Re}\left[\sum_{n \:\in\: {\rm Ex}} \langle e_n| \partial_{i} e_{\rm GS}\rangle \langle \partial_j e_{\rm GS}|e_n\rangle \right ] \;,\\
&=4{\rm Re}\left [\sum_{n} \langle \partial_j e_{\rm GS}|e_n\rangle\langle e_n| \partial_{i} e_{\rm GS}\rangle - \langle \partial_j e_{\rm GS}|e_{\rm GS}\rangle\langle e_{\rm GS}| \partial_{i} e_{\rm GS}\rangle \right ] \;, \\
&= 4{\rm Re}\left [  \langle \partial_j e_{\rm GS}|\partial_{i} e_{\rm GS}\rangle - \langle \partial_j e_{\rm GS}|e_{\rm GS}\rangle\langle e_{\rm GS}| \partial_{i} e_{\rm GS}\rangle\right ] \;,
\end{split}
\end{align}
where in the middle line we extended the sum over all states, and on the last line we used the resolution of the identity $\hat{1}=\sum_n |e_n\rangle\langle e_n|$. The result is precisely that for pure states, Eq.~\eqref{eq:Pure_State_QFI}.

Therefore, we conclude that if the QFIM is singular for the ground state (meaning that the pure-state QFIM determinant computed using the ground state is zero) then the QFIM of the thermal state will also vanish continuously in the $T\to 0$ limit. In most standard cases, even when the QFIM is singular for all pure states, the thermal state QFIM will be non-singular. Our result shows that in such a case raising the temperature will lift the singularity. Therefore metrological sensitivity is \textit{enhanced} by increasing the temperature. This is illustrated in Sec.~\ref{sec:lift_sing}.



\bibliography{bibo}

\end{document}